\documentclass[fleqn,usenatbib]{mnras}

\usepackage{newtxtext,newtxmath}

\usepackage[T1]{fontenc}

\DeclareRobustCommand{\VAN}[3]{#2}
\let\VANthebibliography\thebibliography
\def\thebibliography{\DeclareRobustCommand{\VAN}[3]{##3}\VANthebibliography}

\usepackage{graphicx}	
\usepackage{amsmath}	

\usepackage{enumitem}

\usepackage{xcolor}
\newcommand{\mainred}{\texttt{MAIN-RED}}
\newcommand{\mainblue}{\texttt{MAIN-BLUE}}
\newcommand{\mainbroad}{\texttt{MAIN-BROAD}}
\newcommand{\mwsmain}{\texttt{MWS-MAIN}}
\newcommand{\insitu}{\textit{in situ}}
\newcommand{\ananke}{\texttt{ananke}}
\newcommand{\particleid}{\texttt{ParticleID}}

\DeclareRobustCommand{\msol}{\mathrm{M_{\sun}}}
\DeclareRobustCommand{\rsol}{\mathrm{R_{\sun}}}
\DeclareRobustCommand{\lsol}{\mathrm{L_{\sun}}}
\DeclareRobustCommand{\kpc}{\mathrm{kpc}}
\DeclareRobustCommand{\enbid}{E{\fontsize{6}{10}\selectfont N}B{\fontsize{6}{10}\selectfont I}D}
\DeclareRobustCommand{\galaxyflow}{G{\fontsize{6}{10}\selectfont ALAXY}F{\fontsize{6}{10}\selectfont LOW}}

\title{AuriDESI: Mock Catalogues for the DESI Milky Way Survey}
\author[Kizhuprakkat et al.]{Namitha Kizhuprakkat,$^{1,2}$\thanks{E-mail: namitha96@gapp.nthu.edu.tw} 
Andrew P. Cooper,$^{1,2}$
Alexander H.~Riley,$^{3}$
Sergey E. Koposov,$^{4,5,6}$ 
\newauthor Jessica Nicole~Aguilar,$^{7}$
Steven~Ahlen,$^{8}$
Carlos~Allende Prieto,$^{9,10}$
David~Brooks,$^{11}$
Todd~Claybaugh,$^{7}$
\newauthor Kyle~Dawson,$^{12}$
Axel~de la Macorra,$^{13}$
Peter~Doel,$^{11}$ 
Jaime~E.~Forero-Romero,$^{14,15}$
Carlos~Frenk,$^{3}$
\newauthor Enrique~Gaztañaga,$^{16,17,18}$
Oleg Y. Gnedin,$^{19}$
Robert J.~J.~Grand,$^{20}$
Satya~Gontcho A Gontcho,$^{7}$
\newauthor Klaus~Honscheid,$^{21,22}$
Robert~Kehoe,$^{23}$
Martin~Landriau,$^{7}$
Marc~Manera,$^{24,25}$
Aaron~Meisner,$^{26}$
\newauthor Ramon~Miquel,$^{24,27}$
Jundan~Nie,$^{28}$
Francisco~Prada,$^{29}$
Mehdi~Rezaie,$^{30}$
Graziano~Rossi,$^{31}$
Eusebio~Sanchez,$^{32}$
\newauthor Michael~Schubnell,$^{33}$
Hee-Jong~Seo,$^{34}$
Gregory~Tarlé,$^{33}$
Monica~Valluri$^{19}$ and
Zhimin~Zhou$^{28}$
\\
$^{1}$Institute of Astronomy and Department of Physics, National Tsing Hua University, Hsinchu 30013, Taiwan\\
$^{2}$Center for Informatics and Computation in Astronomy, National Tsing Hua University, Hsinchu 30013, Taiwan\\
\dots\\
\textit{Affiliations are listed at the end of the paper}}

\title{AuriDESI: Mock Catalogues for the DESI Milky Way Survey}
\date{Accepted 2024 June 03; Received 2024 May 30; in original form 2024 January 31}

\pubyear{2024}

\begin{document}
\label{firstpage}
\pagerange{\pageref{firstpage}--\pageref{lastpage}}
\maketitle

\begin{abstract}
The Dark Energy Spectroscopic Instrument Milky Way Survey (DESI MWS) will explore the assembly history of the Milky Way by characterising remnants of ancient dwarf galaxy accretion events and improving constraints on the distribution of dark matter in the outer halo. We present mock catalogues that reproduce the selection criteria of MWS and the format of the final MWS data set. These catalogues can be used to test methods for quantifying the properties of stellar halo substructure and reconstructing the Milky Way's accretion history with the MWS data, including the effects of halo-to-halo variance. The mock catalogues are based on a phase-space kernel expansion technique applied to star particles in the Auriga suite of six high-resolution $\Lambda$CDM magneto-hydrodynamic zoom-in simulations. They include photometric properties (and associated errors) used in DESI target selection and the outputs of the MWS spectral analysis pipeline (radial velocity, metallicity, surface gravity, and temperature). They also include information from the underlying simulation, such as the total gravitational potential and information on the progenitors of accreted halo stars. We discuss how the subset of halo stars observable by MWS in these simulations corresponds to their true content and properties. These mock Milky Ways have rich accretion histories, resulting in 
a large number of substructures that span the whole stellar halo out to large distances and have substantial overlap in the space of orbital energy and angular momentum. 
\end{abstract}





\section{Introduction}

The Dark Energy Spectroscopic Instrument Milky Way Survey (DESI-MWS) will observe nearly 7 million stars in our galaxy using the DESI multi-object spectrograph  installed on the Mayall 4-m telescope at Kitt Peak National Observatory \citep{2022APC}. MWS will be carried out in bright sky conditions that are not used for DESI's concurrent cosmological redshift surveys \citep{Schlegel2015,DESI2016,Parker2018,Martini2018,Levi2019}.  The footprint of DESI covers approximately 14,000 sq.\ deg.\ on the sky, of which about 9800 sq.\ deg.\ are in the Northern Galactic Cap and 4400 sq.\ deg.\ in the Southern Galactic Cap \citep{DESI2016,Schlegel2015,Parker2018,Martini2018,Levi2019}.  Over 5 years from 2021, MWS will observe this footprint to a uniform (extinction-corrected) limiting magnitude of 19 in the DESI Legacy Imaging Survey $r$ band. When complete, MWS will be the largest wide-area spectroscopic survey of stars to this depth. The data will provide radial velocities and chemical abundances, primarily for stars in the thick disc and stellar halo. When combined with Gaia's astrometry, such data can constrain the history of accretion of progenitor dwarf galaxies into the Milky Way halo and the present-day distribution of Galactic dark matter. The selection function of MWS, described in  \ref{sec:mws_photometric_cat}, is simple and inclusive, which will simplify  statistical interpretation and forward modelling of the data. A comprehensive overview of MWS is given in \citet{2022APC}.

The classical description of the stellar halo as a single `smooth' component of galactic structure \citep[e.g.][]{EggenBellSandage1962,Chiba2000} has given way to a much more complex picture, in large part revealed by large surveys like SDSS/SEGUE \citep{SDSS2006, Yanny2009Segue,Juric2008,Ivezic2008,Connie2009,Connie2022} and {\it Gaia} \citep{Gaia2016,Gaia2018,GaiaEDR32021}. Large-scale observations of the Milky Way broadly support the theoretical predictions of the cold dark matter (CDM) model, in which gravitational clustering drives a succession of mergers between galaxies like the Milky Way and numerous dwarf galaxy `progenitors'. Such progenitors are disrupted due to tidal forces from the dark matter halo and the central galaxy, gradually building up a diffuse, metal-poor accreted component comprising only a tiny fraction of the total stellar mass bound to the central galaxy (\citealt{BKW2001,Bullock&Johnston2005,Cooper2010,Font2011}; for recent reviews see \citealt{Bland-Hawthorn2016,Helmi2018}). This process is inherently stochastic; variations in the masses, infall times, and orbits of the accreted progenitors from one system to another are therefore encoded in the properties of their accreted stellar haloes \citep[e.g.][]{Deason_wechsler2016,
Amorisco2017}.

In this picture, the apparent smoothness of the classical `inner' halo of the Milky Way is attributed, at least in part, to highly phase-mixed debris from a small number of relatively massive accretion events. Most such events are expected to occur early in the history of the galaxy, and hence to result in debris confined to relatively small apocentres. The broken power-law density profile of the stellar halo may arise from the correlated apocentres of such an event \citep{EricBell2008,Deason2013}. 
This picture has received further support from the recent discovery of the `Gaia-Enceladus-Sausage` (GES), a relatively metal-rich, kinematically hot component that appears to dominate the halo to a galactocentric radius of at least 25~kpc \citep{Belokurov2018,Helmi2018,Bonaca2020}. Up to 25~kpc, more than 50~per~cent of halo stars may belong to GES; beyond 30~kpc, where the density of the halo as a whole begins to decline more steeply, the fraction of probable GES stars also decreases \citep{Lancaster2019,Deason2012,AlisDeason2018,Evans2020}. This suggests, in broad terms, that GES may contribute to the previously observed dichotomy in the metallicity and kinematics of  `inner' and `outer' halo stars \citep[e.g.][]{Carollo2007,Deason2011,Battaglia2017,Carollo2021,Nissen&Schuster2010}. 
 One or more `\insitu{}' formation processes may also contribute `smooth' halo components \citep{Cooper2015,perez-villegas2017}, including periods of chaotic, kinematically hot star formation before the growth of the present-day Galactic disc \citep{Belokurov&Kravtsov2022,Bonaca2017}. Robustly distinguishing such \insitu{} components from the accreted stellar halo remains a significant challenge.

The dynamical mixing timescale becomes longer in the outer regions of the Galactic potential, allowing stellar substructure from accretion events to remain coherent in configuration space \citep{DeLucia&Helmi2008,Helmi2008,Zolotov2009,Deason2012b}. Such structures have been observed, including the Sagittarius stream, the Orphan--Chenab stream, and many more, including very thin and coherent streams arising from the disruption of globular clusters \citep{Ibata1995,Juric2008,Watkins2009,mateu2022galstreams}. Several of these known substructures, including the Sagittarius stream, GD1, the Orphan--Chenab stream, and the Virgo overdensity, are at least partly covered by the DESI footprint. The footprint also includes many surviving satellite galaxies, some of which may have undergone moderate tidal stripping in the past \citep{MYWang2017}.

Observations from new spectroscopic surveys like H3 \citep{H32019}, MWS \citep{2022APC}, WEAVE \citep{Weave}, SDSS-V \citep{SDSSV}, 4MOST \citep{4most}, and PFS \citep{PFS2022}, in combination with Gaia, will further improve our understanding of the stellar halo and its substructures. The discovery and characterisation of these structures can reveal significant events in the formation of the Milky Way and hence enable more detailed comparisons with accretion histories expected for similar galaxies in the $\Lambda$CDM cosmogony. The stellar populations of these structures potentially probe the earliest epochs of low-mass galaxy formation, a regime that may not be well represented by surviving satellite galaxies or dwarfs  in the field \citep{Naidu2022}.

Large surveys of stellar populations covering a significant fraction of the virial volume of the galaxy, such as MWS, are expected to be complex and difficult to interpret. 
Empirical synthetic star catalogues, such as the Besançon model \citep{Robin2003}, are the most common means of interpreting large Galactic surveys. These attempt to `fit' to observations of the real Milky Way using a small set of parameterized functions for the density and star formation histories of different galactic components. Predictions can be made by extrapolating these models (for example, to greater depth or different bandpasses). They can be refined and updated by adjusting their parameters or by adding components to match new observations. These empirical models are, by construction, an accurate representation of the known Galaxy. However, they are not forward models and hence have no predictive power concerning the history of galactic accretion events or the interaction between those events and the growth of the central galaxy. 

Forward modelling with ab initio simulations is, therefore, an important, complementary approach to understanding the features present in the stellar halo and their cosmological context.  Such simulations can also help to guide the robust recovery of those features from the data. However, forward models can only be compared to the Milky Way in a statistical sense. Although the range of cosmological initial conditions producing `Milky Way-like' systems can be narrowed down for greater efficiency and relevance, no models produced in this way are likely to  match the Milky Way in all respects. Moreover, techniques for simulating the baryonic processes involved in galaxy formation, such as star formation and feedback, remain highly uncertain; two cosmological models of a Milky Way analogue, run from identical initial conditions but using different state-of-the-art codes, are unlikely to produce the same star formation histories for the central galaxy and all its satellites. This may be the dominant source of uncertainty in current models of galactic stellar haloes \citep{Cooper2017}.

To distinguish between alternative predictions and interpret new observations, it is essential to compare existing models to the available data as directly and realistically as possible. In addition to the selection of suitably realistic initial conditions (Milky Way analogues) and physical recipes, the production of synthetic (`mock') catalogues is an essential part of such comparisons. This technique was pioneered in the context of observational cosmology, in which mock catalogues of the cosmic galaxy population, produced by post-processing N-body simulations of the large-scale structure, have proved invaluable for connecting observational statistics (for example, the galaxy correlation function) to fundamental theoretical predictions. In the context of the Milky Way, the Galaxia code \citep{Sharma2011} has provided, in addition to synthetic star catalogues derived from  empirical models Milky Way, a means of producing such catalogues from \citet{Bullock&Johnston2005} N-body simulations of dwarf galaxy accretion. This was achieved with a phase-space sampling method based on the EnBiD phase-space volume estimator \citep{Sharma&Steinmetz2006}.

 This method was developed further by \citet{Lowing2015}, who produced synthetic star catalogues from the Aquarius suite of cosmological $N$-body simulations \citep{Springel2008,Cooper2010}. The \citet{Lowing2015} method assigns a clipped Gaussian `expansion kernel' to each stellar $N$-body particle, into which individual stars are scattered, corresponding to the fraction of the associated stellar population visible to a Solar observer (see Section~\ref{sec:star_particle_expansion}). Similar techniques have been applied to the Latte/FIRE-2 simulations \citep{Wetzel2016} using the \ananke{} framework \citep{Robyn2020, Nguyen2023, Thob2023}. 
 
 \citet{Grand2018} applied this method to Auriga \citep{Grand2017}, a suite of cosmological zoom simulations of 30 Milky Way mass galaxies (described in more detail in Section~\ref{AuriGaia Simulations}). The Auriga galaxies, which are mostly disc-dominated, have sizes, star formation histories, and rotation curves spanning ranges characteristic of $\mathrm{L}_\star$ galaxies like the Milky Way. Auriga is one of the largest and highest-resolution sets of such simulations currently available. \citet{Grand2018} constructed mock catalogues based on Auriga, called the AuriGaia ICC-MOCKS\footnote{Alongside the ICC-MOCKS, an alternative method using the SNAPDRAGONS code \citep{Hunt2015}, referred to as HITS-MOCKS in \citet{Grand2018}, was used to create an alternative set of AuriGaia mocks. In this paper, we only use the ICC-MOCKS version of AuriGaia.} using the method of \citet{Lowing2015}. These aimed to reproduce the content and error model of the Gaia DR2 dataset (see Section \ref{AuriGaia Simulations}). 
 The Auriga galaxies, and hence the AuriGaia mocks, span a range of plausible realizations of Milky Way-like systems useful for exploring the capability and biases of a DESI-like survey, and for developing algorithms to detect substructure, distinguish different accretion histories and infer properties of the gravitational potential from stellar kinematics. 
 
In this paper, we introduce AuriDESI, a new suite of mock catalogues that provide mock realizations of the DESI MWS dataset based on the Auriga galaxies. To produce AuriDESI, we have augmented AuriGaia with Legacy Survey photometry, applied the MWS target selection criteria, and added mock measurements that correspond to those made by the  MWS reduction pipelines, with empirical error models derived from the DESI Early Data Release (EDR) \citep{2022APC, Koposov2024}. These mock catalogues enable direct comparisons between MWS and the Auriga simulations.

\begin{figure*}
    \centering
    \includegraphics[width=15cm]{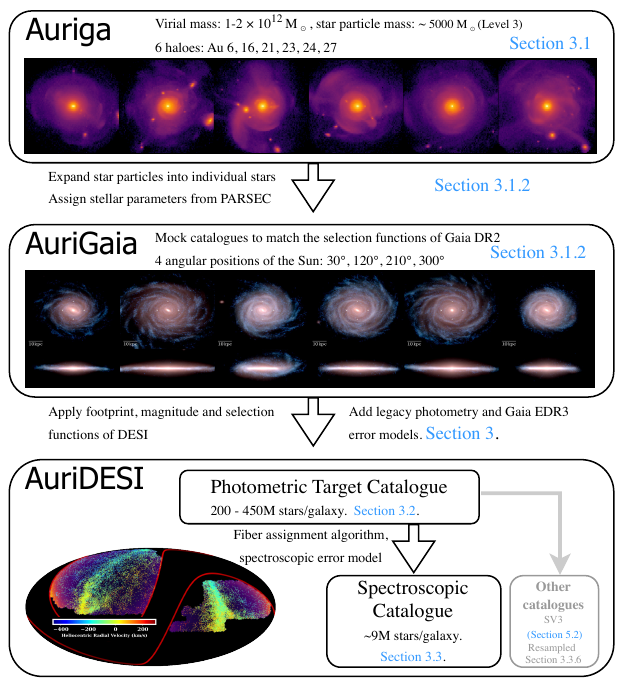}
    \caption{Schematic summary of the datasets discussed in this paper and their relationship to one another, with the corresponding sections indicated. Our work is based on the Auriga simulations (top panel) and the AuriGaia mock catalogues (middle panel). Here we present AuriDESI (bottom panel) which comprises, for each Auriga galaxy, a mock photometric target catalogue (corresponding to stars that DESI is able to observe) and a mock spectroscopic catalogue (a realization of the final DESI MWS dataset, including an error model for spectroscopic observables such as radial velocity and metallicity). In practice we provide four sets of catalogues for each galaxy, corresponding to the different Solar positions used for the AuriGaia catalogues. We also provide supplementary catalogues to address the MWS SV3 dataset and differences in the relative density of different DESI target classes between Auriga and the real Milky Way (`resampled' spectroscopic catalogues).}
    \label{fig:flowchart}
\end{figure*}

Figure \ref{fig:flowchart} gives a schematic summary of the various mock datasets discussed in this paper. Like all spectroscopic surveys, MWS comprises both a `parent' (or `input') catalogue of photometrically selected \textit{potential targets} (obtained by applying selection criteria to sources from the DESI Legacy Imaging survey) and a \textit{spectroscopic} catalogue of  measurements for the subset of stars in the parent catalogue that are actually observed by the DESI spectrograph over the course of the survey \citep[see the following section and][]{myers22a}. For each Auriga galaxy, we apply comparable photometric cuts to produce a mock MWS photometric target catalogue, and then select a subset of those targets using the DESI fiber assignment algorithms to produce a mock MWS spectroscopic catalogue. The spectroscopic catalogues reproduce the error model for the fundamental spectroscopic observables of MWS, such as radial velocity and metallicity.
 
We proceed as follows. In Section \ref{sec:mws_data_products}, we briefly describe the footprint and selection functions used in MWS, and the photometric and spectroscopic catalogues produced by the survey. In Section \ref{sec:auridesi_mocks}, we review the Auriga simulations, the existing AuriGaia mocks, and the new AuriDESI photometric and spectroscopic mock catalogues. In Section \ref{sec:auriga_as_seen_by_desi}, we explore the haloes from a "simulator's perspective" to guide future interpretations of the mock catalogues. We focus on the bulk properties of the stellar halo the dominant progenitors of the accreted debris visible to DESI. Section \ref{sec:applications} describes applications of the AuriDESI mocks, including forecasts for the complete Milky Way survey, validation of the mocks by comparison to the DESI Survey Validation (SV) dataset, and a brief exploration of how surviving satellites are represented in the mocks. We discuss our results and conclude in Section~\ref{discussion}.

We make the AuriDESI mock catalogues publicly available at at \url{https://data.desi.lbl.gov/public/papers/mws/auridesi/v1}. The data model is described in Appendix \ref{mock data model}.

\section{DESI MWS data products}
\label{sec:mws_data_products}

Before describing the mock catalogues themselves, we briefly summarize the real DESI MWS data products and note differences with the mock catalogues we provide. A complete description of the MWS design and observing strategy is given in \citet{2022APC}.

\begin{table}
    \begin{tabular}{l|l|r}
    Category & Description &  MWS spectra  \\
    \hline
    \hline
    \multicolumn{3}{l}{\textit{Highest priority (sparse; $16<r<19$)}}  \vspace{0.1cm}\\
    \texttt{MWS-BHB} & Blue horizontal branch & 17,706 ($55\%$)  \\
    \hline
    \multicolumn{3}{l}{\textit{Baseline priority (core MWS halo sample; $16<r<19$)}}\vspace{0.1cm} \\
    \mainblue{} & Metal-poor turnoff, giants & 3,693,518 ($32\%$) \\
    \mainred{}  & Distant halo giants & 805,794 ($31\%$) \\
    \hline
    \multicolumn{3}{l}{\textit{Lower priority (high $\pi$ or $\mu$, or without Gaia astrometry; $16<r<19$})}\vspace{0.1cm} \\
    \mainbroad{}  & Disk dwarfs (some giants)  & 2,077,222 ($19\%$) \\
    \hline
    \multicolumn{3}{l}{\textit{Lowest priority (fiber-filling, $19<r<20$)}}\vspace{0.1cm} \\
    \texttt{FAINT-BLUE} & Metal-poor turnoff, giants & 4,606,314 ($7\%$) \\
    \texttt{FAINT-RED} & Distant halo giants & 960,134 ($11\%$) \\
    \end{tabular}
    \caption{Simplified summary of those MWS target categories that are included in the AuriDESI mock catalogues, with the (approximate) number of MWS spectra in the final MWS dataset and the corresponding spectroscopic completeness \citep{2022APC}. Note that this does not include all target categories in the DESI-MWS survey, for example the sparse, high-priority \texttt{MWS-WD}, \texttt{MWS-RRLYR} and \texttt{MWS-NEARBY} samples. Categories are ordered by the priority with which they are assigned fibers for a given DESI observation, as indicated. See text, appendix~\ref{name:mws selection criteria} and \citet{2022APC} for further details.}
    \label{tab:targets}
\end{table}

\subsection{MWS photometric target catalogue}
\label{sec:mws_photometric_cat}

MWS is expected to yield spectroscopic observations for nearly 7 million stars over the course of the five-year DESI survey. These stars are drawn from a pool of potential targets (a parent sample) constructed by applying astrometric and photometric selection criteria to the DESI Legacy Imaging Survey DR9 source catalogue, combined with Gaia EDR3 astrometry \citep{myers22a}. In the case of MWS, the resulting photometric \textit{target catalogue} comprises $30.4$~million sources. Sources in this photometric target catalogue are assigned to one or more categories, as described below. Opportunities for observation with a DESI fiber are then allocated to a subset of targets, based on priorities associated with the categories to which they belong.

\begin{figure*}
    \centering
    \includegraphics[width=\linewidth]{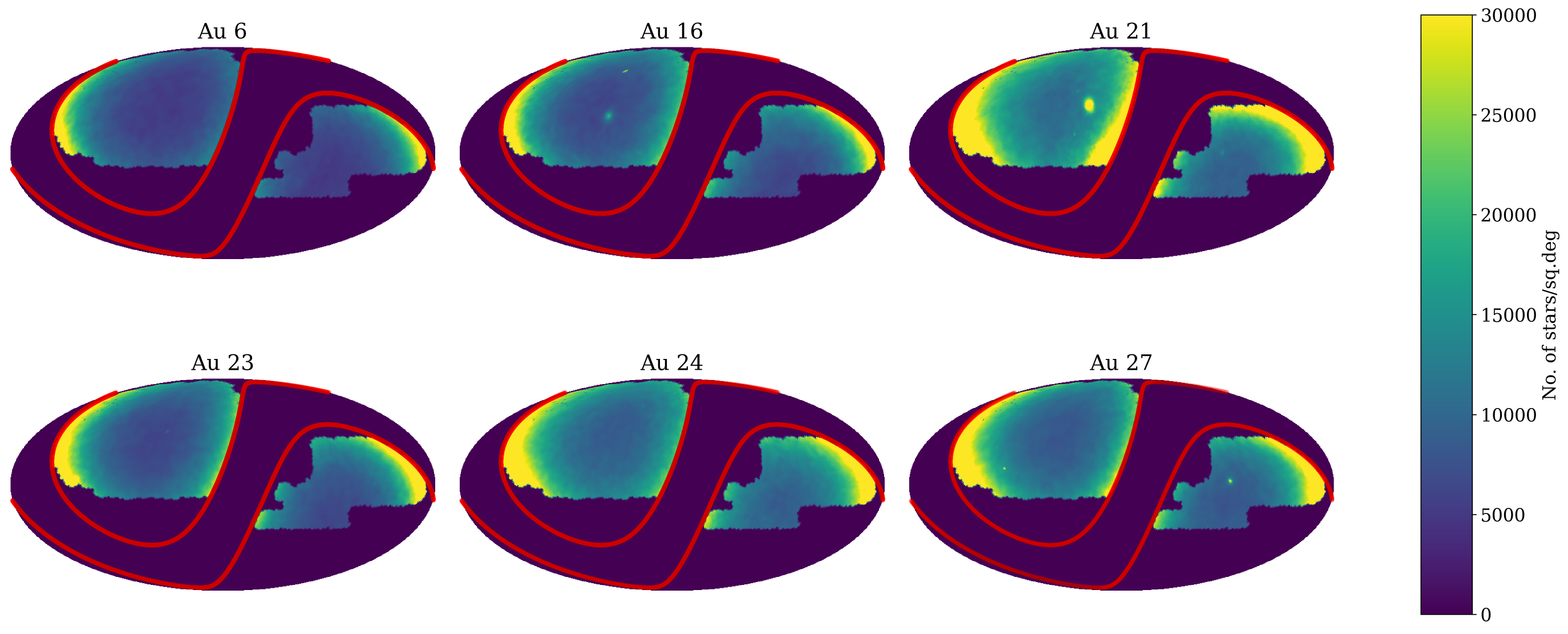}
    \caption{The DESI MWS footprint imposed on the six AuriGaia haloes, in equatorial coordinates, for a fiducial choice of Solar galactic longitude with the Sun's position fixed at a galactocentric distance of 8 kpc and a height of 20 pc above the galactic midplane. The colour scale shows the number of AuriDESI mock stars (those meeting the MWS selection criteria) per square degree. Localized patches of high density correspond to massive satellites, most prominent in Au 16, Au 21, and Au 27. The two red solid lines represent the galactic latitude limits of the DESI survey at $|b| = \pm 20 ^{\circ}$.}
    \label{fig:skyplot}
\end{figure*}

\subsubsection{Survey footprint}
\label{sec:mws_footprint}

The tiling strategy used to observe the DESI survey footprint is customised for the dark and bright time science programs (defined by sky brightness, seeing and other survey efficiency metrics measured in real time). MWS shares the focal plane with the low-redshift Bright Galaxy Survey \citep[BGS;][]{Hahn2022}) under bright sky conditions. The entire footprint is divided into 5675 tiles for the bright-time program, arranged into four overlapping `passes' with around 1400 tiles in each pass. More details on the DESI tiling strategy are given in (Raichoor et al. (in preparation)). 

\subsubsection{Main MWS sample} 
\label{sec:mws_main_sample}

The mock catalogues we describe in this paper address only the so-called MWS \textit{main sample}, which comprises almost all the 7 million stars to be observed by MWS. Main sample targets are divided into three mutually exclusive categories, \mainblue{}, \mainred{}, and \mainbroad{}, based on their colour, parallax ($\pi$), and proper motion ($\mu$). The union of these three categories includes all stars in the magnitude range $16 < r < 19$, where $r$ is the LS $r$-band magnitude after correcting for dust extinction. In Table~\ref{tab:targets} we provide a simplified summary of these subsets and other target categories relevant to the mock catalogues in this paper.

Briefly, \mainblue{} comprises all stars with colour $g-r<0.7$, with no requirements on their astrometry. Redder stars are selected either into the \mainred{} or \mainbroad{} samples. \mainred{} stars must meet additional astrometric requirements designed to favour distant halo giants, whereas \mainbroad{} stars either fail those astrometric cuts (indicating they are likely disk dwarfs) or do not have well-measured Gaia astrometry. \mainbroad{} is given lower fiber assignment priority than \mainblue{} and \mainred{}. Within the DESI footprint, the spectroscopic completeness is expected to be $\sim30$~per~cent for the \mainblue{} and \mainred{} samples and $\sim20$~per~cent  for the \mainbroad{} sample (completeness will vary with latitude). In addition, all DESI observations (including the dark time surveys) include spectrophotometric standards, with colours and magnitudes corresponding to the metal-poor main sequence turnoff. Standard stars are almost all \mainblue{} targets as well, but they have a higher probability of observation over the course of the survey because a minimum number of standards must be observed in each DESI field. This effect is included in the assignment of fibers to stars in the mock catalogue. Further details of the main sample target selection criteria, including the specific astrometric cuts that separate \mainbroad{}  and \mainred{}, are given in Appendix \ref{name:mws selection criteria}. 

\subsubsection{Other MWS target categories} 
\label{sec:mws_other_targets}

MWS will include several other target categories, as described in \citet{2022APC}. These mostly comprise rare but scientifically valuable types of star. Of these additional categories, only \texttt{MWS-BHB} and \texttt{MAIN-FAINT} are included in the AuriDESI mock catalogues described here (see Section~\ref{sec:bhb_targets} for further details of the BHB selection). Among the target categories included in the mock catalogues, BHBs are given the highest priority for fiber assignment, yielding an estimated spectroscopic completeness of $\gtrsim50\%$. \texttt{MAIN-FAINT} extends the \mainblue{} and \mainred{} selection to the magnitude range $19 < r < 20$ (after extinction correction). There is no faint equivalent of \mainbroad{}; $r>19$ stars having Gaia astrometry consistent with disk dwarfs, or no astrometric measurements, are not targeted. \texttt{MAIN-FAINT} targets have the lowest fiber assignment priority of all MWS categories, hence their spectroscopic completeness is $\lesssim10\%$.

MWS will observe other categories which, like \texttt{MWS-BHB}, comprise relatively small numbers of rare stars and have higher fiber assignment priority than the main sample. These target categories are not currently included in the AuriDESI mock catalogues; they include stars within 100 pc of the Sun (\texttt{MWS-NEARBY}), RR Lyrae variables, and a highly complete sample of white dwarfs to the Gaia magnitude limit. Finally, the DESI surveys include several ``secondary" science programs, which use either dedicated pointings of the telescope or spare fibers in regular survey pointings that have no available primary science target. None of these secondary programs are included in AuriDESI.

\subsection{MWS spectroscopic catalogue}
\label{sec:mws_spectroscopic_cat}

The DESI spectra are processed by the Redrock code (Bailey et al. (in preparation)). Redrock is optimised for redshift-fitting and classifying extragalactic spectra, rather than measuring stellar radial velocities and atmospheric parameters at the accuracy needed for the MWS science cases. MWS has developed three dedicated pipelines that extend the data products provided by Redrock. As described in Section~\ref{spectroscopic catalogue}, the mock spectroscopic catalogues we present here correspond to results from one of these pipelines, RVS, which measures radial velocities and other stellar atmospheric parameters ([Fe/H], $\log\,g$, [$\alpha$/Fe], $V \sin i$), along with their uncertainties\footnote{We do not generate $V \sin i$ data for the mock stars.}. RVS is based on the algorithm for fitting to spectral templates introduced in \citet{Koposov2011}, specifically the publicly available python implementation, RVSpecfit \citep{Koposov2019}. 

We currently do not produce mock observables for the MWS SP pipeline, an alternative, more detailed analysis of atmospheric parameters and individual chemical abundances based on the FERRE code \citep{CarlosAllende2006} or the MWS WD pipeline, which is tailored to the measurement of parameters for white dwarfs. More details of RVS and the other pipelines are given in \citep{2022APC}, along with results of their validation against the DESI early data release.

\begin{table*}
    \begin{tabular}{c|c|c|c|c|c|c|c|c|c|c|c|}
       \hline
       Halo & $\frac{M_{\rm{vir}}}{(10^{12}\,\msol)}$ & $\frac{M_*}{(10^{10}\,\msol)}$ & $\frac{R_d}{(\rm{kpc})}$ & $\frac{V_c (\rsol)}{(\rm{km\,s^{-1})}}$ & $\frac{M_{*,\rm{acc}}}{(10^{9}\,\msol)}$ & $\frac{M_{*,\rm{top 10}}}{(10^{9}\,\msol)}$ & $\frac{M_{*,\rm{top 10,DESI}}}{(10^{9}\,\msol)}$ &
       $\frac{f_{\rm{blue}}}{\%}$ & $\frac{f_{\rm{red}}}{\%}$ & $\frac{f_{\rm{broad}}}{\%}$ &
       $\frac{f_{\rm{ins,>10kpc}}}{\%}$\\
       \hline
       Au 6 & 1.01 & 6.1 & 3.3 & 224.7 & 7.86 & 7.66 & 1.40 & 71 & 26 & 2.6 & 18.03 \\ 
       Au 16 & 1.50 & 7.9 & 6.0 & 217.5 & 9.34 & 8.67 & 1.81 & 71 & 27 & 2.5 & 38.01\\ 
       Au 21 & 1.42 & 8.2 & 3.3 & 231.7 & 19.37 & 19.06 & 3.54 & 75 & 23 & 2.1 & 26.74 \\
       Au 23 & 1.50 & 8.3 & 5.3 & 240.0 & 14.35 & 13.71 & 2.20  & 71 & 25 & 3.9 & 33.25\\
       Au 24 & 1.47 & 7.8 & 6.1 & 219.2 & 12.54 & 11.89 & 1.84  & 76 & 28 & 2.3 & 34.97\\
       Au 27 & 1.70 & 9.5 & 3.2 & 254.5 & 15.40 & 14.38 & 2.54  & 72 & 25 & 3.7 & 23.66\\
       \hline
       MW  & $1.3\pm0.3$ & $6\pm1$ & $2.6\pm0.5$ & $238\pm15$ \\
       \hline
    \end{tabular}
    \caption{Properties of the AuriGaia and AuriDESI mock catalogues. Columns: (1) halo number, (2) virial mass, (3) stellar mass within the virial radius (excluding satellites), (4) disc scale length, (5) circular rotation velocity at 8~kpc from the Galactic centre, (6) accreted stellar mass, (7) total stellar mass of the top ten most massive progenitors in the accreted halo visible to DESI, (8) stellar mass from the top ten most massive progenitors visible to DESI (excluding satellites), (9-11) ratio of mass of \mainblue{}, \mainred{} and \mainbroad{} targets, respectively, to the total stellar halo mass visible to DESI, (12) ratio of \insitu{} stellar mass outside 10 kpc to the total halo mass. The last row gives values for the Milky Way. Columns (1) to (5) are taken from \protect\citet{Grand2018} and values for the Milky Way are taken from \protect\citet{Bland-Hawthorn2016}. The remaining columns are obtained from the mock catalogues described in this paper.}
    \label{tab:table1}
\end{table*}

\section{AuriDESI Mock Catalogues}
\label{sec:auridesi_mocks}

The following sections present in detail the data and processing steps we use to create mock MWS datasets, as illustrated in Fig~\ref{fig:flowchart}. Specifically, we describe:

\begin{itemize}[leftmargin=1.5em,labelwidth=1em,labelindent=0em,itemindent=!]
    \item The Auriga simulations and AuriGaia mock catalogues (\ref{AuriGaia Simulations}):
        \begin{itemize}
            \item The similarities and differences between the Auriga galaxies and the Milky Way (\ref{sec:auriga_is_not_the_mw});
        \item The expansion of star particles into individual stars (\ref{sec:star_particle_expansion}).
    \end{itemize}
    \item The AuriDESI photometric target catalogues (\ref{sec:photometric_target_catalogues}): 
    \begin{itemize}
         \item Our updates to include LS photometry and the Gaia EDR3 error model in AuriGaia (\ref{sec:aurigaia_updates});
         \item Comparison of the AuriDESI photometric target catalogue to the fiducial Galaxia model used to benchmark the real MWS target selection (\ref{sec:footprint_and_targets});
        \item The definitions and labels involved in separating stars formed in situ in the host galaxy from those accreted from progenitor satellites (\ref{sec:satellite_labels}).
    \end{itemize}
    \item The construction the AuriDESI spectroscopic catalogues (\ref{spectroscopic catalogue}):
    \begin{itemize}
            \item The DESI fiber assignment algorithm (\ref{sec:fiber_assignment});
            \item MWS targets not included in AuriDESI (\ref{sec:not_included});
            \item Our treatment of BHB targets  (\ref{sec:bhb_targets});
            \item Our empirical error model for spectroscopic observables, based on early DESI observations (\ref{sec:spectro_errors}).
    \end{itemize}

\end{itemize}

As shown in Fig.~\ref{fig:flowchart}, alongside the AuriDESI mock photometric catalogues and spectroscopic catalogues described in this section, we also provide a set of `resampled spectroscopic mocks' (\ref{sec:resampled_mocks}), which address a specific issue with the relative density of different types of MWS targets in Auriga and are not used for any of the analysis in this paper, and a set of `SV3' mocks (\ref{sec:sv3}),which select a subset of targets that correspond the DESI Early Data Release.

\subsection{Auriga and the AuriGaia mock catalogues}
\label{AuriGaia Simulations}

Our AuriDESI mocks are built on the AuriGaia ICC-MOCKS (hereafter referred to as ``AuriGaia"), which in turn were derived from the Auriga simulation suite, comprising cosmological magnetohydrodynamical zoom simulations of Milky Way analogue dark matter haloes\footnote{The complete Auriga simulation (snapshot) data are publicly available, as described in \citet{Grand2024}.}\citep{Grand2017, Grand2018}. All the Auriga haloes were chosen to be isolated at the present day (redshift $z=0$) and to have virial \footnote{Virial mass is defined as the mass enclosed by a sphere with mean matter density 200 times the critical density, $\rho_\mathrm{crit} = 3H^2(z)$/(8$\pi$G).} masses in the range $1$ -- $2 \times 10 ^{12}\,\msol$. Table~\ref{tab:table1} summarizes the properties of the six haloes for which AuriGaia mocks were produced. These are the Auriga simulations available at resolution level 3, which has a dark matter particle mass $\simeq4 \times 10^4\,\msol$ and a baryonic (star) particle mass $\simeq5 \times 10^3\,\msol$ \citep{Grand2018}. 

AuriGaia provides four mock catalogues for each halo, differing in the angular position of the solar observer in the plane of the galactic disc (30$^{\circ}, 120^{\circ}, 210^{\circ}, 300^{\circ}$) with respect to the bar major axis. In all cases, the Sun is taken to be at a galactocentric radius of 8 kpc and a height of 20 pc above the galactic mid-plane. More details on how the solar position is fixed with respect to the galactic centre are given in \citet{Grand2018}. As shown in Figure 2 of \citet{Grand2018}, the young thin disc scale heights in the six haloes at the solar radius range from 302 to $\sim$ 430 pc, and the thick disc scale heights range from 1103 to 1436 pc. The vertical structure of the Auriga discs near the Sun is therefore broadly comparable to the Milky Way \citep[thin and thick disc scale heights of $\sim300\pm50$~pc and $900\pm180$~pc respectively;][]{Bland-Hawthorn2016}. Although we provide AuriDESI catalogues for all the AuriGaia solar positions, the analyses in this paper refer to the 30$^{\circ}$ version unless otherwise noted; we show later that many bulk properties of a MWS-like sample are not particularly sensitive to the choice of angle.

Two variants of AuriGaia catalogue are available, with alternative treatments of dust extinction: one assuming an empirical Milky Way dust distribution and another which does not include any dust extinction. We use the latter as the base for AuriDESI. This gives the user flexibility to impose a dust model of their choice. Because the MWS $16 < r < 19$ selection criterion is based on extinction-corrected magnitudes, it is not necessary to impose a dust model to select a MWS-like sample (although this neglects any uncertainty in the extinction correction and possible incompleteness in regions of very heavy obscuration). 

\subsubsection{Similarities and differences between Auriga and the Milky Way}
\label{sec:auriga_is_not_the_mw}

Of the six Auriga galaxies used here, Au 6 is the closest Milky Way analogue, based on the stellar mass within its virial radius, its star formation rate, and its morphology. The Au6 disc scale length is 3.3~kpc, comparable to that of the Milky Way \citep[$\sim$2.6~kpc;][]{Bland-Hawthorn2016}. Au 16 and Au 24 have larger discs, with scale lengths of $\sim 6$ kpc. Au 21, Au 23, and Au 27 have prominent ongoing interactions with massive satellites (we discuss this further below). 

The Auriga stellar haloes are more massive and metal-rich than the stellar halo of the real Milky Way, according to current observational estimates. Several factors contribute to this discrepancy. First, the different galaxies have a range of accretion histories, and an average virial mass slightly larger than that of the Milky Way. Second, as noted by \cite{Monachesi2019}, the Auriga galaxies appear more similar to the Milky Way observations when the accreted component is considered in isolation, implying that the mismatch is due in part to an extended, massive in-situ halo component that forms in Auriga\footnote{Our definition of the \insitu{} halo includes stars that form in streams of cold gas stripped from accreting satellites, because this gas is associated with the central potential at the time of star formation.}. Third, the subgrid star formation and feedback models in Auriga result in dwarf galaxies (including the progenitors of the stellar halo and the surviving satellites) that have stellar masses considerably higher than the average for their pre-accretion halo masses obtained from abundance matching \citep{BWC2013,Simpson2018,Monachesi2019}. Moreover, \cite{Grand2021} used higher-resolution simulations to show that Auriga satellite galaxies with luminosities $L_V \gtrsim 10^5 \, \lsol$ are about 0.5 dex more metal rich at fixed luminosity \citep[see Figure 13 in][]{Grand2021}, relative to the observed mass-stellar metallicity relation \citep[e.g.][]{Kirkby2013,Amorisco2017}. This apparent excess in both mass and metallicity  also contributes to the discrepancy between the Auriga stellar haloes and that of the Milky Way \citep[e.g.][]{Deason_wechsler2016,Monachesi2019}.

These differences mean that our mock catalogues are not "fine-tuned" replicas of the real Milky Way in the manner of empirical models such as Galaxia. In many cases, the Auriga galaxies may be sufficiently Milky Way-like that differences between the total mass and metallicity of the Milky Way's stellar halo and the mocks will be of secondary importance, although this will depend on the specific application. To some extent, such effects can be explored by comparing the six Auriga galaxies to one another. It may be possible to "adjust" the mocks for consistency with the Milky Way (for example, by sub-sampling a fraction of stars to better match the density profile, or by re-scaling metallicities using observed mass-metallicity relations). Although such post hoc adjustments could provide insight into sampling effects, they risk breaking the self-consistency between the mix of halo stellar populations, orbits, accretion histories and other properties of the host galaxy, which is arguably the main advantage of using mock catalogues based on forward models. We therefore present the mock catalogues without any such adjustments.

\subsubsection{Star particle expansion in AuriGaia}
\label{sec:star_particle_expansion}

Simulations such as Auriga contain \textit{star particles}: tracers representing single-age, single-metallicity stellar populations (SSPs). Making a mock catalogue involves `expanding' each of these massive star particles into a large number of individual \textit{mock stars} (the number being determined by the total mass of the star particle). 

As described in \citet{Grand2018}, the AuriGaia mocks were created using the phase-space kernel-sampling technique of \citet{Lowing2015}. Using a Chabrier IMF (as in the Auriga simulations), the present-day mass function of the constituent stars associated with each $\sim5000\,\msol$ star particle is sampled in mass intervals of 0.08--120 $\msol$. Using isochrones, these individual stars are assigned atmospheric parameters according to their initial mass. The stars are then distributed over the phase-space volume associated with their parent star particle using a 6D Gaussian smoothing kernel. The extent and orientation of a star particle's kernel in each dimension is based on a measure of distance from its neighbours in phase space, computed separately for star particles associated with each progenitor galaxy. This multidimensional smoothing aims to preserve correlations between the positions and velocities of stars in the original simulation\footnote{For example, where the tidal disruption of a progenitor has created a long, thin stream of star particles, the aim is to preferentially `interpolate' mock stars along the length of the stream, preserving any velocity gradient, without thickening it or artificially inflating its velocity dispersion. Treating star particles from each progenitor separately avoids `cross talk' between overlapping streams.}.

AuriGaia adopts the PARSEC model isochrones (release v1.2S)\footnote{Specifically, we use the isochrones as provided by CMD v3.0: \url{http://stev.oapd.inaf.it/cgi-bin/cmd}}. These isochrones are sampled with a metallicity grid spanning 0.0001 $\leq Z \leq 0.06$ and ages 6.63 $\leq \log_{10} (\mathrm{age/yr}) \leq$ 10.13 \citep{Bressan2012,Chen2014,Chen2015,Tang2014,Grand2018}. Star particles outside these ranges of metallicity and age are matched to the nearest isochrone. The isochrones therefore do not accurately represent stellar populations with $\mathrm{Z}_* < 0.0001$, equivalent to $\mathrm{[M/H]} \approx \mathrm{[Fe/H]} \leq -2.2$. We adopt the default treatment of post-AGB and horizontal branch stars in these isochrones. Of significance for our mock BHB sample, predictions for horizontal branch stars assume a constant mass loss on the red giant branch corresponding to a Reimers parameter of 0.2 \citep{Reimers1975,Bressan2012}. Although the resulting stellar halo horizontal branch has broadly similar magnitude to observations, its morphology  differs in detail, in particular having a less pronounced blue hook and lacking the gap associated with the instability strip. As noted above, we do not include white dwarf stars in our mock catalogue, and the isochrones do not account for the photometric (or spectroscopic) effects of binaries. Finally, potentially significant uncertainties related to the treatment of very low-mass stars in the v1.2S PARSEC isochrones are described in Section~\ref{sec:footprint_and_targets} and discussed further in Appendix \ref{isochrones}.

In this paper, we consider both the mock stars in our AuriDESI catalogues and the properties of the underlying distribution of simulated star particles, taking care to distinguish between the two. Mock stars can be matched to star particles using the unique identifier for star particles, \particleid{}, which we provide in our catalogues. Every mock star expanded from a given star particle has the same \particleid{}. Hereafter, we refer to mock stars simply as `stars', except where it is necessary to distinguish them from stars in the real DESI survey. We refer explicitly to `star particles' from the simulation where necessary.

\subsection{AuriDESI Photometric Target Catalogues}
\label{sec:photometric_target_catalogues}

We construct AuriDESI photometric target catalogues for each AuriGaia mock by selecting stars that meet the MWS selection criteria (section \ref{sec:mws_photometric_cat} and appendix~\ref{name:mws selection criteria}) and fall within the DESI footprint, defined by the union of the set of DESI survey tiles. In this section we compare the mock photometric target catalogues to the fiducial Galaxia Milky Way model used to benchmark the real MWS photometric target catalogue in \citet{2022APC}.
 
\subsubsection{Updated photometry and astrometry}
\label{sec:aurigaia_updates}

The MWS selection functions are magnitude dependent and are based on the Legacy Survey photometric system. The AuriGaia catalogues provide G, BP, RP, and UBVRI photometry. We therefore supplement AuriGaia with mock photometry in this system, by interpolation of the PARSEC isochrones using the initial masses of stars from AuriGaia. The MWS selection function also uses Gaia photometry and astrometric measurements from Gaia EDR3.  We have therefore updated the error models in AuriGaia from DR2 to DR3, using PyGaia\footnote{\url{https://github.com/agabrown/PyGaia}}.  In DR3, uncertainties are reduced relative to DR2 by $\sim20\%$ in parallax and by a factor of 2 in proper motion.

\subsubsection{Comparison to a fiducial Milky Way model}
\label{sec:footprint_and_targets}

Fig. \ref{fig:skyplot} shows the sky distribution of stars in the AuriDESI target catalogues. In all cases, as in the real MWS photometric target catalogue, the stellar density increases towards lower galactic latitudes at the edges of the footprint.
The compact overdensities visible at large galactocentric distance in Au 16, Au 21, and Au 27 correspond to surviving satellites.

We have compared the basic properties of these mock surveys to a fiducial mock catalogue based on the Galaxia model \citep{Sharma2011}. As in \citet{2022APC}, we modified the power-law density profile of the Galaxia stellar halo to introduce a break, which improves the agreement with recent constraints on the observed surface density of stars at distances > 20~kpc. \citep{Watkins2009,Deason2011,Sesar2011}. Since Galaxia is an empirical model of the Milky Way, this comparison also illustrates, approximately, the large-scale differences between AuriDESI and the real MWS photometric target catalogue. 

\begin{figure}
    \centering
    \includegraphics[width=\linewidth]{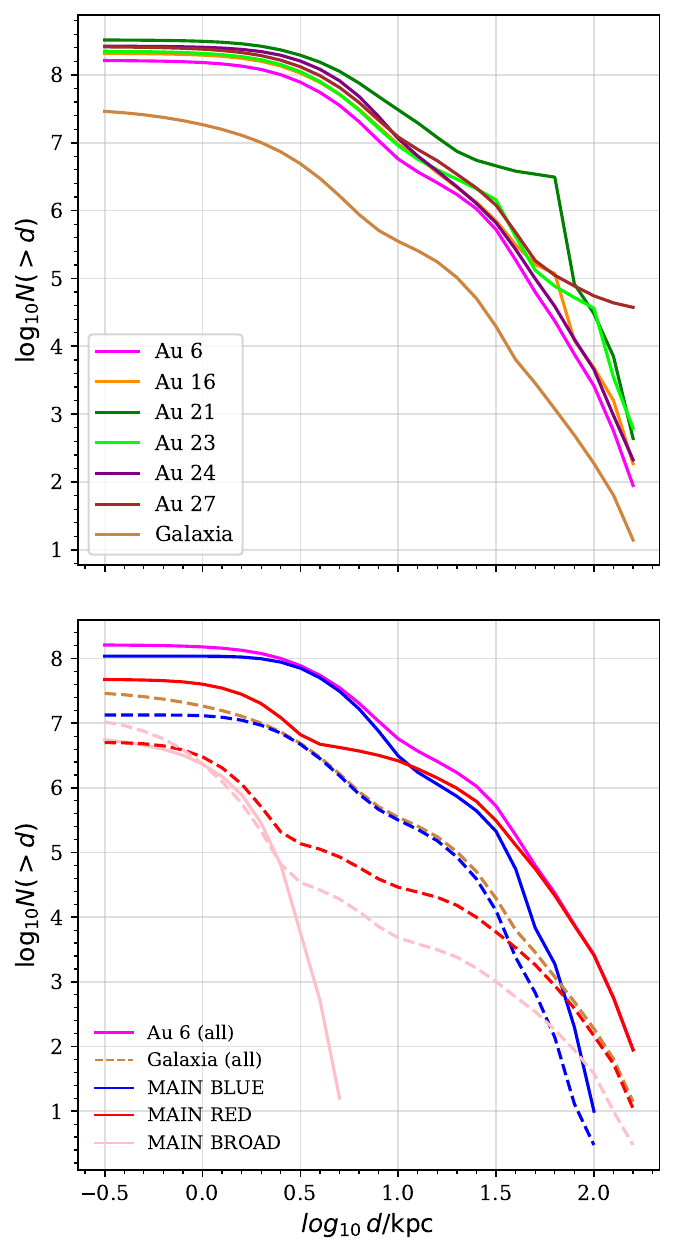}
    \caption{Distribution of heliocentric distances, $d$, for stars in mock MWS surveys(restricted to $d> 300$ pc). The top panel shows the distributions for stars in the six AuriGaia haloes, after imposing the MWS footprint and selection criteria. The light brown line shows the distribution of stars selected in the same way from a Galaxia model (our broken power-law stellar halo variant, see text). The bottom panel shows the separate distance distributions of the three MWS main survey target classes in AuriGaia Au 6 (solid lines), which can be compared to the equivalent predictions from Galaxia (dashed lines).} 
    \label{fig:dist}
\end{figure}

\begin{figure}
    \centering
    \includegraphics[width=\linewidth]{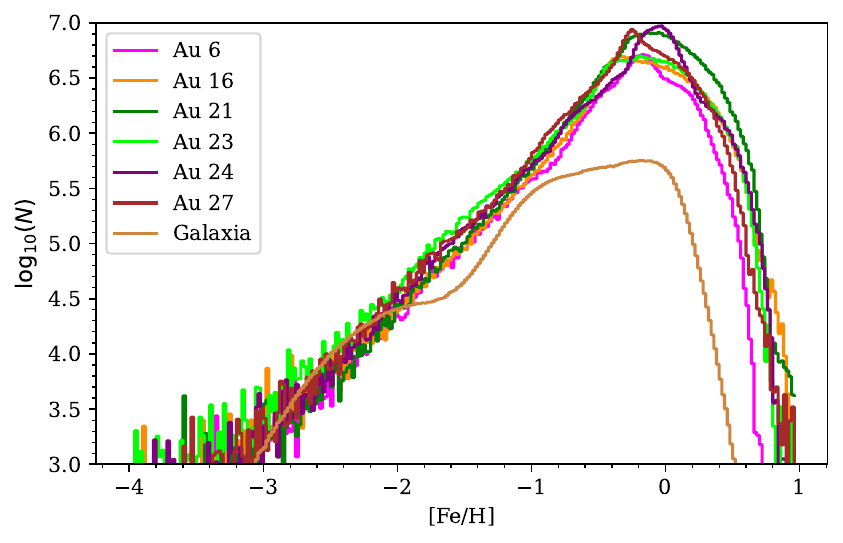}
    \caption{Metallicity distributions for stars in the six AuriDESI mock photometric target catalogues, compared with that of our Galaxia model (light brown).}
    \label{fig:fehhist}
\end{figure}

The top panel of Fig.~\ref{fig:dist} shows the number of stars beyond a given heliocentric distance in each mock catalogue. The distance distribution is broadly similar in each halo up to $\sim30\,\kpc$, except for Au 21. The strong feature in that halo around $60\,\kpc$ corresponds to a particularly massive satellite. Although the overall density distributions of the mock catalogues are broadly similar to that of the fiducial Galaxia model, the mocks have an order of magnitude more stars at almost all radii. This is because, as noted above, the Auriga stellar haloes are more massive compared to the real Milky Way \citep{Monchesi2016}.

The bottom panel of Fig.~\ref{fig:dist} separates the stars in Au 6 into the three different MWS main target classes. Again, we compare these profiles to Galaxia. This shows which target class (and hence which kind of stellar population) dominates the sample at a particular distance. The \mainblue{} selection dominates out to $\sim10\,\kpc$ (dominated by thick disc main-sequence turnoff stars in this region). The \mainblue{} and \mainred{} samples contribute equally in the range $10 < d < 30 \,\kpc$. The \mainred{} sample (predominantly halo giants) dominates at larger distances. To first order, the shape of this distribution is determined by the apparent magnitude range of the MWS selection function.

In the mock photometric target catalogues, the \mainbroad{} sample (dominated by redder main sequence stars in the thin disk) makes only a small contribution to the total counts. This is very different from the fiducial Galaxia model, in which \mainbroad{} stars dominate the sample up to $\sim10\,\kpc$ \citep[see also figure 5 in][]{2022APC}. Although differences in the star formation histories of the galaxies may contribute to this discrepancy, we believe it is mostly an artifact of the different PARSEC model versions used by Galaxia and AuriGaia (hence AuriDESI). With the more recent PARSEC isochrones used by AuriGaia \citep[release v1.2S,][]{Chen2014}, a significant fraction of the fainter, redder part of the thin disc main sequence, which makes a large contribution to \mainbroad{}, falls outside the MWS apparent magnitude range. Since the number counts in the real DESI photometric target catalogue are in reasonable agreement with those predicted by Galaxia, this argues against the use of the 1.2S PARSEC isochrones for this purpose. We discuss this issue in more detail in Appendix \ref{isochrones}. We intend to improve the treatment of these very low mass stars in future work. However, the MWS science goals focus on the more distant stellar halo, where this effect is less important; for the rest of this paper we concentrate on the more distant stars in the sample. The main effect of this discrepancy between the PARSEC isochrones and observations on the mock photometric target catalogue is an under-estimate of the number of \mainbroad{} stars relative to \mainred{} stars (the effect on the spectroscopic catalogue is discussed in Section~\ref{spectroscopic catalogue}.)

Fig.~\ref{fig:fehhist} shows the metallicity distribution of the stars within the MWS footprint. All the AuriDESI catalogues have peaks at super-solar metallicities and hence are substantially more metal-rich than both the Galaxia model and the real Milky Way \citep{Monchesi2016,RobertGrand2018,Halbesma2020}. As noted in section~\ref{sec:auriga_is_not_the_mw}, the Auriga satellites have higher stellar mass than expected from abundance matching relations, and higher metallicity at fixed stellar mass than expected from the observed mass-metallicity relation. These differences most likely originate from the particular sub-grid models used for star formation and feedback in Auriga. Another possibility is that the virial masses of the Auriga galaxies (which determine the typical \insitu{} star formation history and accreted satellite mass function, and hence the history of chemical enrichment for both the stellar halo and the central galaxy) may not be matched closely enough to the virial mass of the Milky Way. There is also a possibility that the evolutionary history of the Milky Way is sufficiently unusual that such differences would arise even if the dark matter halo mass was well-matched in Auriga, particularly for a small sample of simulations.

\begin{figure*}
    \centering
    \includegraphics[width=\linewidth]{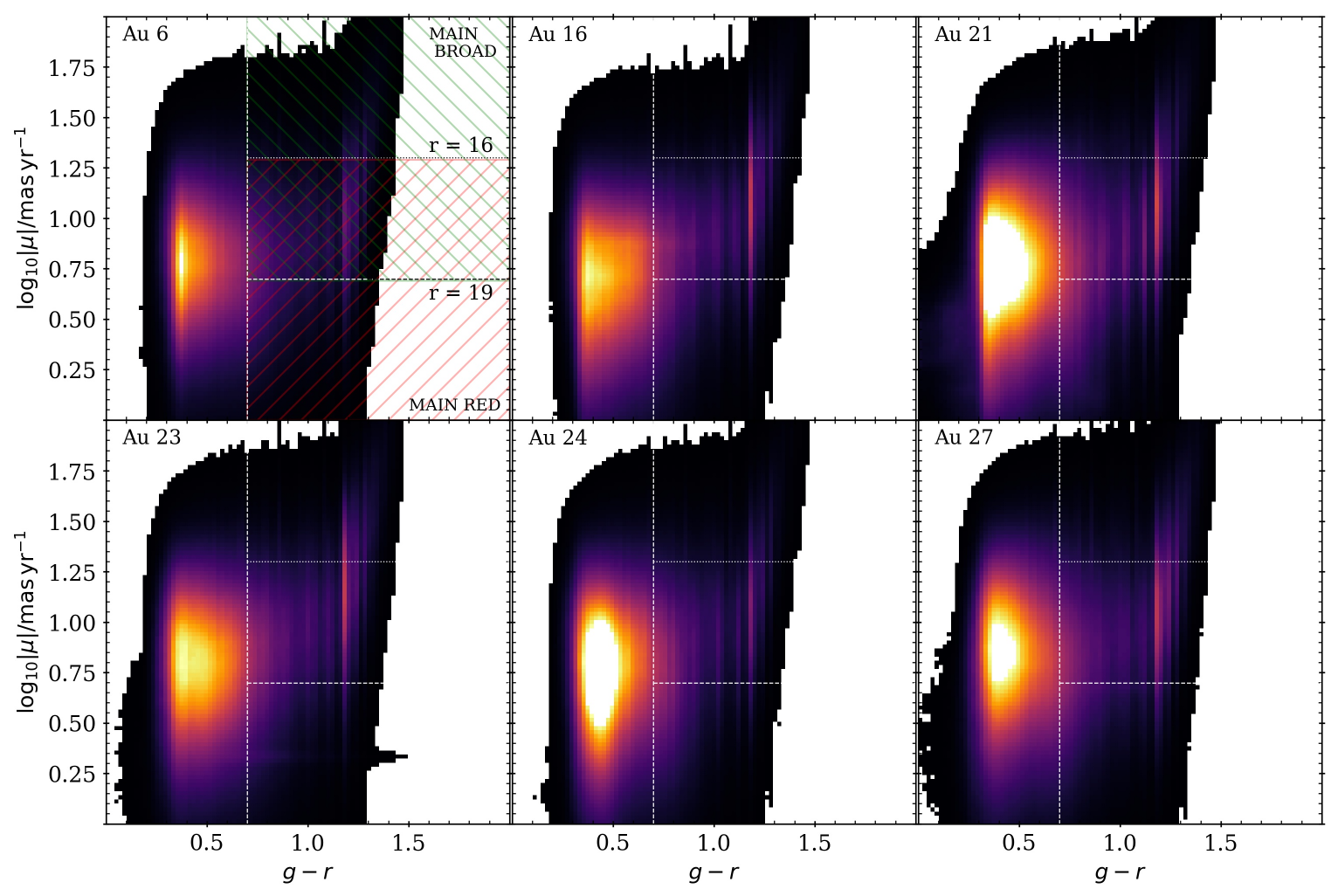}
    \caption{The distribution of AuriDESI targets in the space of proper motion and $g-r$ colour. The vertical line (at $g - r = 0.7$) separates the \mainblue{} sample from the \mainred{} and \mainbroad{} samples. The horizontal lines show how the separation between \mainred{} (red hatched region) and \mainbroad{} (green hatched region) depends on the magnitude of the source over the range of magnitudes covered by the survey, from $r = 16$ (upper line) to $r = 19$ (lower line).}
    \label{fig:pmcolor}
\end{figure*}

Fig.~\ref{fig:pmcolor} shows the distribution of stars in the proper motion--colour space that is used to define the MWS main target classes. There is a very clear peak at blue colour, corresponding to \mainblue{} (the metal-poor stellar halo, with low proper motion). The density of this peak differs greatly between haloes. The redder peak on the diagram, corresponding to thin disc stars in the \mainred{} and \mainbroad{} samples, is substantially weaker compared to the corresponding diagram for the real MWS photometric target catalogue \citep[see fig. 3 in][]{2022APC}. This reflects the absence of the lower disc main sequence in the AuriDESI catalogues, as discussed above; in these diagrams it is made even more apparent by the greater number of metal-poor halo stars in AuriDESI. The discreteness in colour is a result of choosing the nearest initial mass gridpoint for a given star when computing the present-day luminosity distribution of each stellar population, rather than interpolating along the isochrone. 

\begin{figure*}
     \centering
     \includegraphics[width=\linewidth]{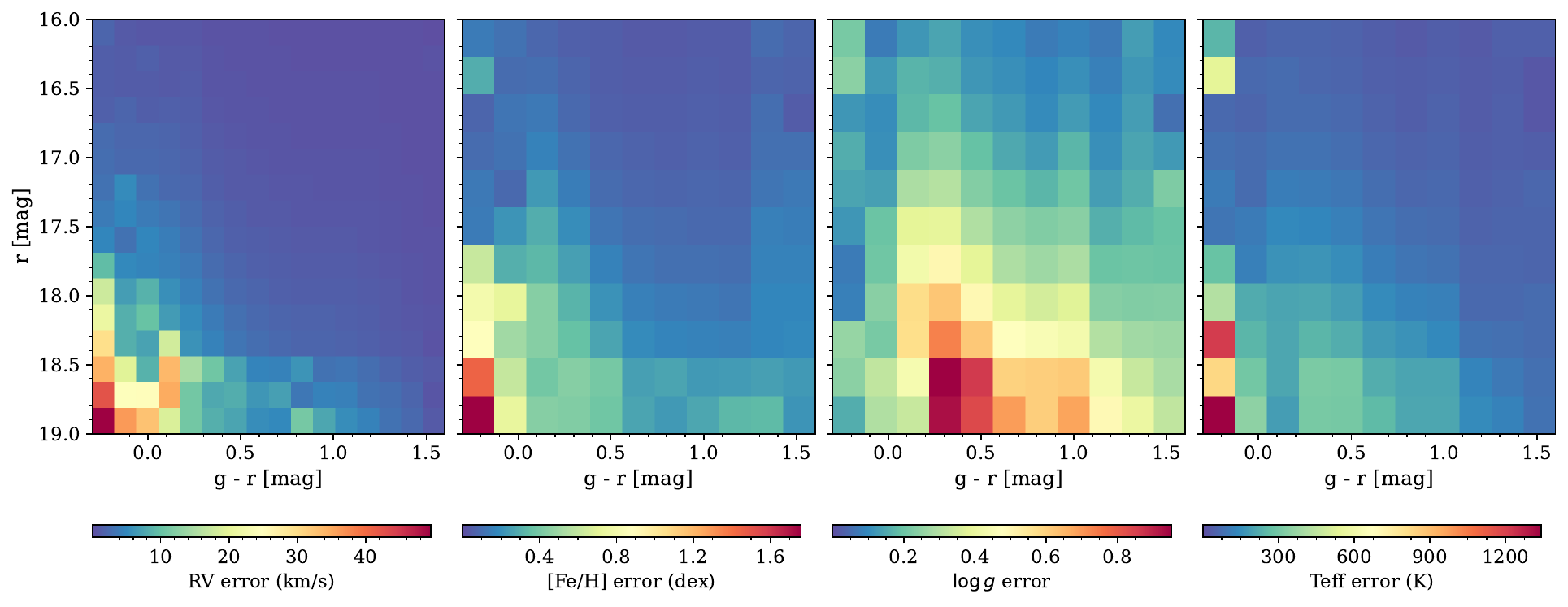}
     \caption{Panels from left to right show the precision of radial velocity, metallicity, surface gravity, and effective temperature respectively, derived from measurements of DESI EDR spectra with the MWS RVS pipeline. We compute errors in these parameters for stars by interpolating their colour and magnitude onto these empirical grids. The limits of the grids are $-0.3 < g - r < 1.8$ and $16 < r < 20$ (extinction corrected magnitudes).}
     \label{fig:desi_err_interp_1}
\end{figure*}

\subsubsection{In situ stars, accreted progenitors and satellites}
\label{sec:satellite_labels}

When studying the mock catalogues, it is useful to know the identify of the gravitational potential to which a star is bound (either the main halo, or one of its surviving satellites) and the identity of the progenitor object that brought a given star into the accreted halo -- much of the analysis in this paper involves associating subsets of halo stars with their progenitors.

Star, gas, and dark matter particles in the Auriga simulations were partitioned into self-bound haloes and subhaloes with the SUBFIND algorithm \citep{Springel2001}. Each simulation contains a `main halo' (the host of the Milky Way analogue) and its satellite subhaloes\footnote{The simulations also contain independent haloes close to the main halo (analogues of the Local Group dwarf galaxies). However, the AuriGaia mocks were only constructed for particles in a sphere of $\sim200~\mathrm{kpc}$ around the main halo, so these more distant neighbours are not included in our mock catalogues.}. Stars have an integer label, \texttt{SubhaloNr}, which indicates the halo to which their parent star particle is associated at the present time. Those with \texttt{SubhaloNr = 0} are bound to the Milky Way analogue (including its accreted and \insitu{} stellar halo, see below); those with \texttt{SubhaloNr > 0} are bound to satellite subhaloes (hence are analogues of the surviving dwarf satellites of the Milky Way, many of which are observed by MWS). Each surviving subhalo is identified by a different positive value of \texttt{SubhaloNr}.

We use merger trees constructed on the Auriga simulation to further partition star particles (and hence stars) bound to the main halo into those formed \insitu{} in the Milky Way analogue and those \textit{accreted} from other \textit{progenitor galaxies} (which, at $z=0$, may be either fully disrupted or intact). The \texttt{TreeID} column provides this information in the mock catalogue: stars with \texttt{TreeID = 0} formed \insitu{}, whereas those with \texttt{TreeID > 0} are accreted. The value of \texttt{TreeID} for accreted stars identifies the progenitor branch of the merger tree in which those stars formed. This label is defined by the direct progenitor branches of the main halo: each of those branches in turn will have many distinct hierarchical progenitors, all of which (in our scheme) will share the same \texttt{TreeID}. 

These direct progenitor branches end when they are tidally stripped to the extent that no self-bound structure can be associated with them in the simulation. For most subhalo branches, this end point occurs long after they have become satellites (i.e. fallen within the virial radius of the main halo). Stars formed after a branch becomes a satellite of the main halo are grouped under the same \texttt{TreeID} as those that formed when the branch was an independent halo, and hence are still counted as `accreted' if they are stripped into the stellar halo of the Milky Way analogue. This simplifies the operational definition of the accreted and \insitu{} stellar halo in our catalogues: \insitu{} star particles (and hence the stars they spawn) are those that were bound to the main halo at the time of their formation; all other stars bound to the main halo at $z=0$ were accreted.

\subsection{AuriDESI spectroscopic catalogues}
\label{spectroscopic catalogue}

To create mock spectroscopic data sets, we apply the full DESI fiber assignment algorithm to each AuriDESI photometric target catalogue. We refer to the subset of stars that are assigned to a fiber -- those that would actually be observed by DESI -- as the AuriDESI spectroscopic catalogue (distinct from the AuriDESI photometric target catalogue described above). Since stars in the spectroscopic catalogue are drawn from the photometric target catalogue, they necessarily have all the same quantities (photometric observations, `true' simulated quantities, and labels) associated with them. For the stars included in the spectroscopic catalogues, we compute additional mock observables corresponding to the measurements made on DESI spectra by the MWS RVS pipeline (see Section~\ref{sec:mws_spectroscopic_cat}).

\subsubsection{Fiber assignment}
\label{sec:fiber_assignment}

The fiber assignment algorithm takes, as input, a list of DESI tiles and targets. The targets are assigned to fibers based on the state of the DESI hardware and predetermined relative priorities for the different target classes. Detailed descriptions of fiber assignment for the bright time program are given in \citet{smith2019}, \citep{Hahn2022} and \citet{EDRvalidation2023, EDRadame2023}. More information on MWS targeting strategies and the expected fiber assignment completeness for the MWS target classes can be found in \citet{2022APC}. In the real MWS, the algorithm assigns fibers to $\sim30$~per~cent of \mainblue{} and \mainred{} targets and $\sim20$~per~cent of \mainbroad{} targets, averaged over the footprint (see Table~\ref{tab:targets}). The spectroscopic completeness is higher for all targets at higher galactic latitudes.

In practice, we apply the DESI fiber assignment algorithm to the union of the AuriDESI photometric target catalogue and the \textit{real} DESI Bright Galaxy Survey target catalogue. This combination accounts for the higher fiber assignment priority of BGS galaxies, which imprints the large-scale structure of the low-redshift galaxy distribution on to the sky distribution of observed MWS targets. We assume the state of the DESI focal plane on 2020/01/01, prior to commissioning of the instrument, in which all fiber positioners are assumed to be working with nominal properties (as opposed to being broken or stuck in position). In this respect, our mock catalogues are somewhat idealised with respect to the final survey dataset, because the operational state of the positioners will evolve over the course of the survey (this may, for example, produce correlations between completeness and the time at which different areas of the sky are surveyed). More detailed modelling of these effects will be included in later updates to the mock catalogues. We assign fibers to targets over all the tiles in the full five-year, four-pass bright time survey; the algorithm accounts for the completion of galaxy observations and hence the greater availability of fibers for stellar targets on later survey passes.

\subsubsection{Spectroscopic catalog data model}
\label{sec:spectro_data_model}

As described in Appendix \ref{mock data model}, the data file for each mock spectroscopic catalogue comprises five FITS extensions: \texttt{RVTAB} (\ref{tab:RVTAB}), which includes information on the measured parameters of stars, including radial velocity, metallicity and effective temperature, and their uncertainties \texttt{FIBERMAP} (\ref{tab:Fibermap}), which includes the targeting data for the stars; \texttt{GAIA} (\ref{tab:Gaia}), which contains mock Gaia observables; \texttt{TRUE\_VALUE} (\ref{tab:true}); and \texttt{PROGENITORS} (\ref{tab:Columns in PROGENITOR extension in AuriDESI}). The latter two extensions provide additional information based on the star particles from the original Auriga simulation, including their provenance as described later in this section.

\subsubsection{MWS targets not included AuriDESI}
\label{sec:not_included}

As noted previously, when assigning spectroscopic fibers, a relatively small number MWS targets with low density but high scientific value (such as white dwarfs\footnote{White dwarfs are given higher priority than BGS galaxies.}, BHBs, RR-Lyraes, and stars within 100 pc of the Sun) are given higher priority than \mwsmain{} stars, to ensure higher completeness. Furthermore, metal-poor F-type stars are selected as spectroscopic standards, which may receive several observations over the course of the survey and hence higher completeness than other \mwsmain{} targets. The white dwarf and 100pc samples are not included in our mock catalogues; the small number of fibers that would otherwise be used for white dwarfs are allocated to BGS targets, whereas those that would be assigned to the 100pc sample are instead assigned to the main MWS target classes.

\subsubsection{BHB targets}
\label{sec:bhb_targets}

For AuriDESI, we use a BHB selection that reproduces the intent of the corresponding MWS BHB selection, but does not use the empirical BHB selection criteria described \citet{2022APC}. This is because the PARSEC isochrones use a simple model for the horizontal branch, which does not correspond in all respects to the locus of observed  BHB stars in the Milky Way. The empirical MWS BHB selection is fine-tuned to the observed locus. In addition, the real BHB sample may be further contaminated by QSOs and blue stragglers. We cannot reproduce this aspect of the selection in the mock catalogues at present, because we do not include those sources of contamination, or the infrared photometry used to identify them.

We therefore employ a different, idealized BHB selection based on the (true, not mock-observed) values of surface gravity ($2.2 < \log\, g < 3.5$) and effective temperature ($5500 < T_\mathrm{eff} < 12000$) in the mock catalogue. This picks out all stars along the horizontal branch of the PARSEC isochrones -- the assumption being that the real MWS BHB selection would pick out stars on the real BHB locus efficiently enough to provide a near-complete sample. As in the real MWS, the stars we identify as BHBs are given higher fiber priority than other MWS targets. We do not explicitly identify RR Lyrae stars in the mocks, although these could be selected as subset of the mock horizontal branch in a similar way to the BHBs. Since  horizontal branch stars are important halo tracers, a more detailed treatment is a priority for future improvement of the mock catalogues.

\subsubsection{Errors for spectroscopic observables} 
\label{sec:spectro_errors}

We obtain empirical error models for spectroscopic observables using data from the DESI survey validation (SV) program, carried out from November 2019 to May 2021. The aim of SV was to understand the quality of the data and to verify that the target selection algorithms and analysis pipelines met requirements for a range of scientific goals. This was done in three stages (SV1, SV2 and SV3). SV3, also called the one percent survey, observed targets in a superset of the final DESI target selection function in a small number of densely sampled fields covering a total of 100 sq. deg.. The DESI early data release (EDR) includes spectra from all three SV programs \citep{2022APC, EDRvalidation2023, EDRadame2023}. 

From the SV3 data, we obtain the median error of radial velocity, metallicity, surface gravity, and effective temperature measurements in bins of colour and magnitude. Fig. \ref{fig:desi_err_interp_1} shows these empirical median errors interpolated smoothly across the colour-magnitude diagram. As discussed in \citet{2022APC}, the MWS RVS pipeline delivers radial velocities accurate to $\simeq 1\,\mathrm{km\, s^{-1}}$ for a large fraction of the sample, with relatively higher radial velocity errors ($\approx10\,\mathrm{km\, s^{-1}}$) towards bluer colours and fainter magnitudes. A similar trend is visible in the other parameters. Although SV3 contains only a small fraction of the number of stars expected in the final main survey sample, it is highly complete and has good coverage of the final selection function; we therefore expect these distributions to be representative of the final sample (although future improvements to the Redrock and RVS pipelines are also expected, see \citealt{2022APC}). For every star, we obtain deviations in each observed quantity by mapping the star to a colour--magnitude bin and treating the corresponding median empirical error as the width of a Gaussian distribution, from which we draw randomly. The sum of the `true' value of the observable and the random perturbation is stored as the `measured' value of the observable in our mock catalogue. The `true' value of the observable is also stored in the \texttt{TRUE\_VALUE} extension (see Appendix \ref{tab:true}). 

\subsubsection{Resampled spectroscopic mock catalogues}
\label{sec:resampled_mocks}

In Section~\ref{sec:footprint_and_targets}, we noted that the photometric target catalogue contains far fewer \mainbroad{} stars than the real MWS photometric target catalogue. This is the result of a feature introduced in the PARSEC isochrones, starting from the 1.2S version which we adopt \citep[][see also appendix~\ref{isochrones}]{Tang2014,Chen2014}. However, in the spectroscopic catalogues, this effect is masked by the fact that the Auriga halos produce many more \mainred{} and \mainblue{} targets than exist in the real Milky Way. The greater density of these high priority targets means that \mainbroad{} stars would not be many assigned fibers, even if their number in the mock photometric target catalogues was similar to that of the real Milky Way. The \mainbroad{} targets are simply swamped by \mainred{} and \mainblue{} targets.

To allow users to assess these sampling issues for themselves, at least in a simplistic way, we also provide a `re-sampled' version of the spectroscopic mock catalogue which, by construction, has the total count for each of \mainred{}, \mainblue{}, \mainbroad{}, \texttt{FAINT-RED}, \texttt{FAINT-BLUE} and \texttt{MWS-BHB} as the real MWS spectroscopic catalogue. To do this, we draw the required number of stars for each category randomly from the AuriDESI photometric target catalogues, and compute mock spectroscopic observables for them. These resampled spectroscopic catalogues may be useful in cases where a `realistic' count of particular target types is required, but they should be used with caution -- by design, they break the correspondence between the chemistry and structure of the Auriga halos and the number of stars that appear in the mock survey. None of the results shown in this paper use these resampled spectroscopic catalogues.

\section{The Auriga galaxies as seen by AuriDESI} 
\label{sec:auriga_as_seen_by_desi}

In this section, we use the AuriDESI photometric target catalogues (containing all stars that meet MWS selection criteria and fall within the DESI footprint) to show that the MWS selection yields a representative sample of the bulk of the stellar halo in the Auriga galaxies. The consequences of sampling only a fraction of these photometric target catalogues with spectroscopic observations will be explored in subsequent sections.

The main issue that we address here is that incomplete sky coverage and finite depth mean that MWS (or any similar survey) has only a partial view of the accreted stellar halo. We can use the mock catalogues to asses the extent to which the stars included in the MWS selection function are representative of the halo as a whole. Moreover, an important objective of galactic archaeology in the Milky Way is to quantify the masses and accretion times of its most significant progenitors. We can therefore ask the same question for each of the most significant progenitors in Auriga -- what fraction of the debris from each progenitor does the mock MWS observe, and how representative are those stars of the progenitor as a whole? These questions are closely related because (at least in Auriga) a handful of the most massive progenitors account for the vast majority of the total mass of the accreted halo \citep{Monachesi2019}.

\subsection{The composition of the accreted stellar halo}
\label{sec:top_10_composition}

For simplicity, we focus on the 10 progenitors that contribute the most stellar mass to each AuriDESI survey\footnote{Ranked according only to the stars selected as mock MWS targets. It makes little difference in practice if we consider the top 10 most massive progenitors in the simulation overall, rather than defining the `top 10' according only to the stars in the mock survey. This choice does not change which progenitors are the `top 10' in almost all cases; in a few cases it changes the rank order of individual progenitors by one or two places.}. These are the debris systems that would be found to dominate a survey like MWS, given sufficient data and an accurate means of distinguishing the orbital and chemical signatures of different progenitors. In the AuriDESI mock catalogues, stars associated with a given progenitor can be identified using their TreeID label.

\begin{figure}
    \centering
    \includegraphics[width=\linewidth]{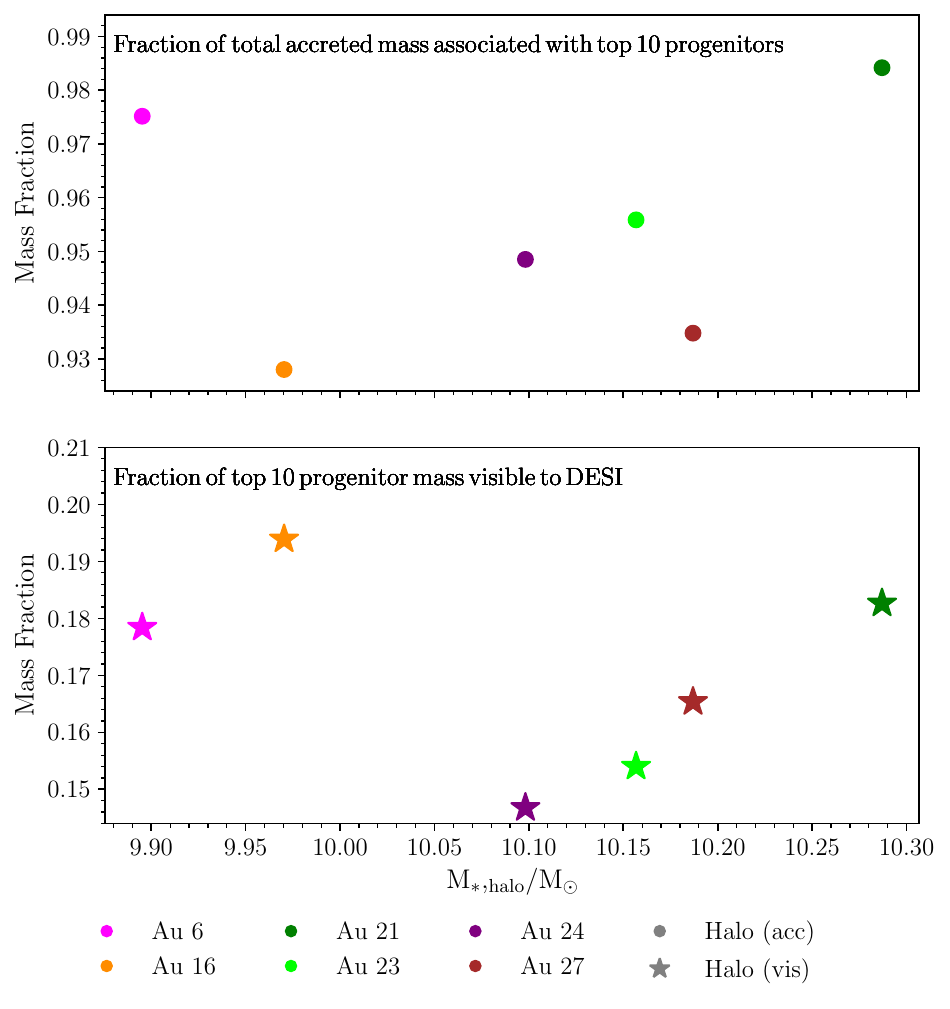}
    \caption{Top panel: points show, for each Auriga halo, the fraction of the total accreted stellar halo mass (shown on the horizontal axis) that is associated with its 10 most massive progenitor galaxies. Bottom panel: the fraction of mass associated with the 10 most massive progenitors (i.e. the quantity reported in the top panel) that is visible with an MWS-like survey footprint and selection function. For example, the 10 most massive progenitors in Au 6 make up $\sim 97\%$ of accreted stellar halo mass. Of that 97\%, $\sim$ 18\% is visible to a DESI-like survey. Symbol colours correspond to different Auriga galaxies as shown in the legend.}
    \label{fig:mass_frac}
\end{figure}

 \begin{figure*}
    \centering
    \includegraphics[width=\linewidth]{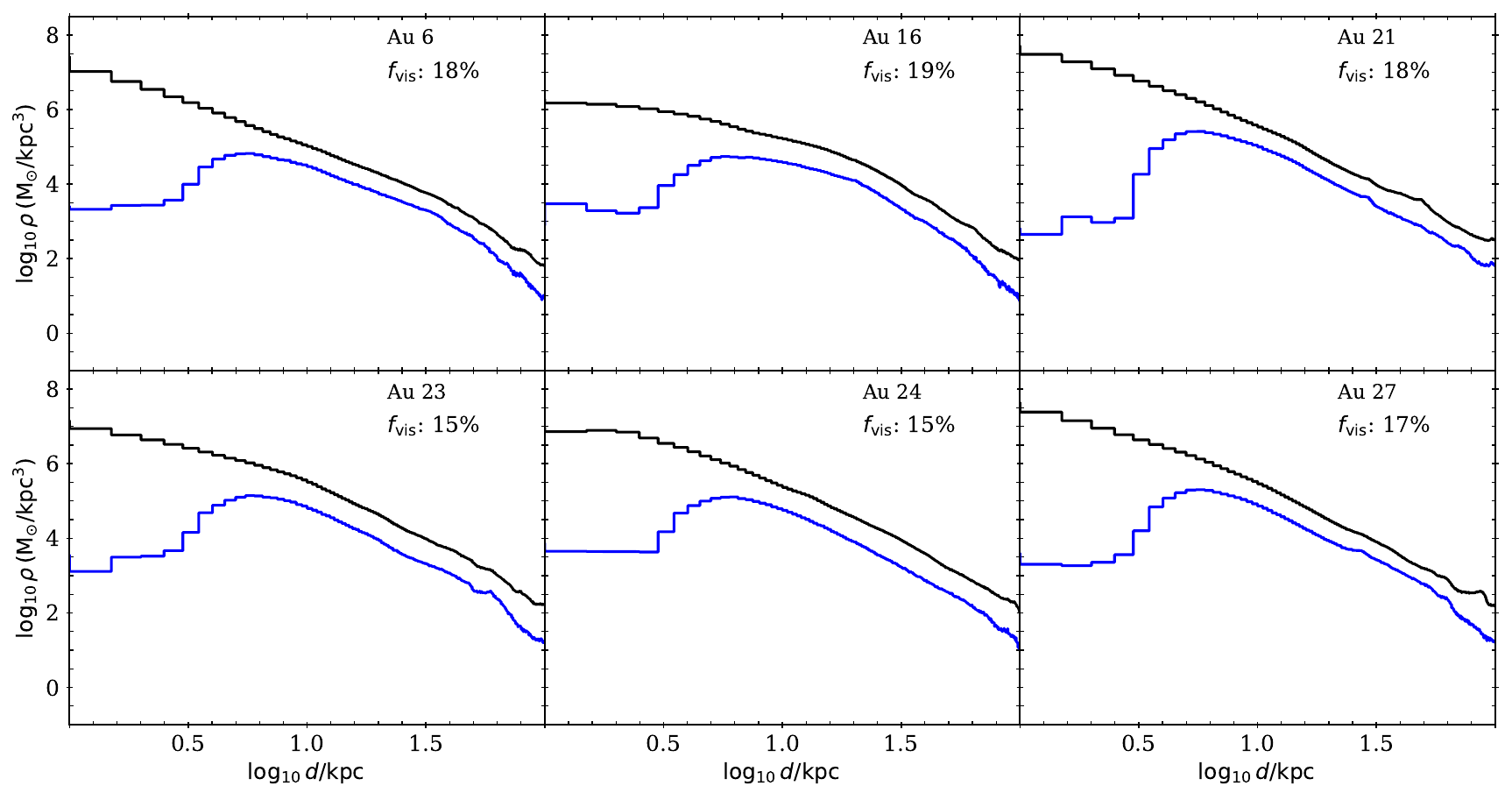}
    \caption{Stellar mass density profiles of star particles in the Auriga simulations. In each panel, the black line represents the density profile of all the accreted star particles and the blue line corresponds to only those star particles that contribute at least one star to the mock catalogue (i.e. to star particles visible within the MWS footprint and selection function; see text). The $f_\mathrm{vis}$ value quoted in each panel is the fraction of mass visible to DESI. Au 16 and Au 21 show strong bumps around 50 kpc, which correspond to massive substructures. The distance is galactocentric; since the DESI survey is conducted from the Solar position and avoids low latitude sky, the centre of the galaxy is not visible.}
    \label{fig:density_snap}
\end{figure*}

The upper panel of Fig.~\ref{fig:mass_frac} shows that, across the six Auriga galaxies, the combined mass of the top 10 progenitors ranges from 93 to 99\% of the total mass of the accreted halo. The lower panel shows the fraction of the combined stellar mass associated with these 10 progenitors that is visible to DESI, i.e. that falls within the MWS footprint and selection function (the definition of this quantity is not trivial, see below). We find this to be approximately 15 to 20\% of the total in each galaxy. Since the top 10 progenitors account for most of the mass of the halo overall, an MWS-like survey in the Auriga galaxies has access to $\gtrsim 15\%$ of the total mass of the stellar halo.

There is an important caveat associated with our definition of the stellar mass that is `visible to DESI'. In general, only some fraction of the total mass of each star particle from the original simulation will be associated with stars in the mock catalogue. For example, a star particle at a distance of 100~kpc will be represented in the mock by only a handful of its brightest giants; most of the mass of the particle will correspond to main sequence stars fainter than the limiting magnitude of MWS. The sampling of the underlying distribution of stellar mass is never complete (even near to the observer, faint stellar remnants and brown dwarfs will not be included) and becomes more stochastic at larger distances. Of course, this effect has to be taken into account to infer stellar density from real observations: for example, the underlying total stellar mass might be estimated from counts of a particular bright tracer, such as K-giants or BHB stars.

Since our intention here is only to provide a broad overview of the difference between the mock catalogues and the full simulations, we assume that the total mass of a simulation star particle is \textit{fully} represented in the mock catalogue if even just one of the stars it spawns is included. We consider star particles as either `visible to DESI' in their entirety (one or more stars in the mock) or not at all (zero stars in the mock). When we quote the mass fraction of a progenitor visible to DESI, as in Fig.~\ref{fig:mass_frac} and the other figures in this section, we use this all-or-nothing measure based on the full masses of the simulation star particles, as opposed to the the masses of the individual stars in the mock catalogue\footnote{This approach is less sensitive to the stochastic sampling of giants at large distances and technical choices in the construction of the mocks (e.g.\ how the density distribution is smoothed by the phase space kernel).}.

Fig.~\ref{fig:density_snap} shows the (galactocentric) stellar mass density profiles of accreted star particles (black lines) in the six haloes. The density profiles of accreted star particles visible to DESI (following the definition above) are shown in blue. 
Beyond $\sim10$~kpc, the density profile of stars visible to DESI follows that of the total mass up to $\sim30$ to $50$~kpc (although an order of magnitude lower in amplitude), and in most cases reasonably closely to $\sim100$~kpc. The stellar mass sampled by the MWS selection is therefore broadly representative of the bulk density structure of the stellar halo. The larger differences at galactocentric distances $\lesssim 5$~kpc are dominated by the restriction of the MWS footprint to high latitudes, $|b| > 20^{\circ}$, which excludes most of the mass in the thin disc (some of which, in these models, is accreted), as well as the innermost halo and bulge. The different accretion histories of the six galaxies are most apparent in the different slopes of their profiles beyond $30$~kpc and the presence of unmixed substructure, visible as small bumps in the profiles at larger distances. We study how debris from individual progenitors contributes to these profiles in the next section.

\subsection{Au 6 in detail}
\label{Halo 6} 

To illustrate the content and application of the AuriDESI mock catalogues, we now examine one Auriga halo in detail. We use Au 6, the closest analogue of the Milky Way based on properties of the central galaxy (see Section \ref{AuriGaia Simulations}). 
The mass of accreted stars in this galaxy, $7.9\times10^{9}\,\msol$, is the lowest among the six Auriga simulations, although somewhat higher than the conventional $\sim10^9\,\msol$ estimate for the Milky Way. The last major merger in Au 6 occurred at a look-back time of 8~Gyr, comparable to estimates of the time at which the GSE progenitor is thought to have merged into the Milky Way \citep{Naidu2021,Grand2018}. However, \citet{Fattahi2019}, in their study of the radial velocity anisotropy, metallicity and accretion time of progenitors across all the Auriga haloes (using lower resolution simulation suite), concluded that Au 6 does not have any kinematic substructure matching their criteria for a close GSE analog (see also table 3 in \citet{Grand2024}).

Throughout this section, we again refer to the 10 most massive progenitors of the accreted stellar halo as defined above. We begin by examining the mock photometric target catalogue, as in the previous section.

\begin{figure*}
    \centering
    \includegraphics[width=\linewidth]{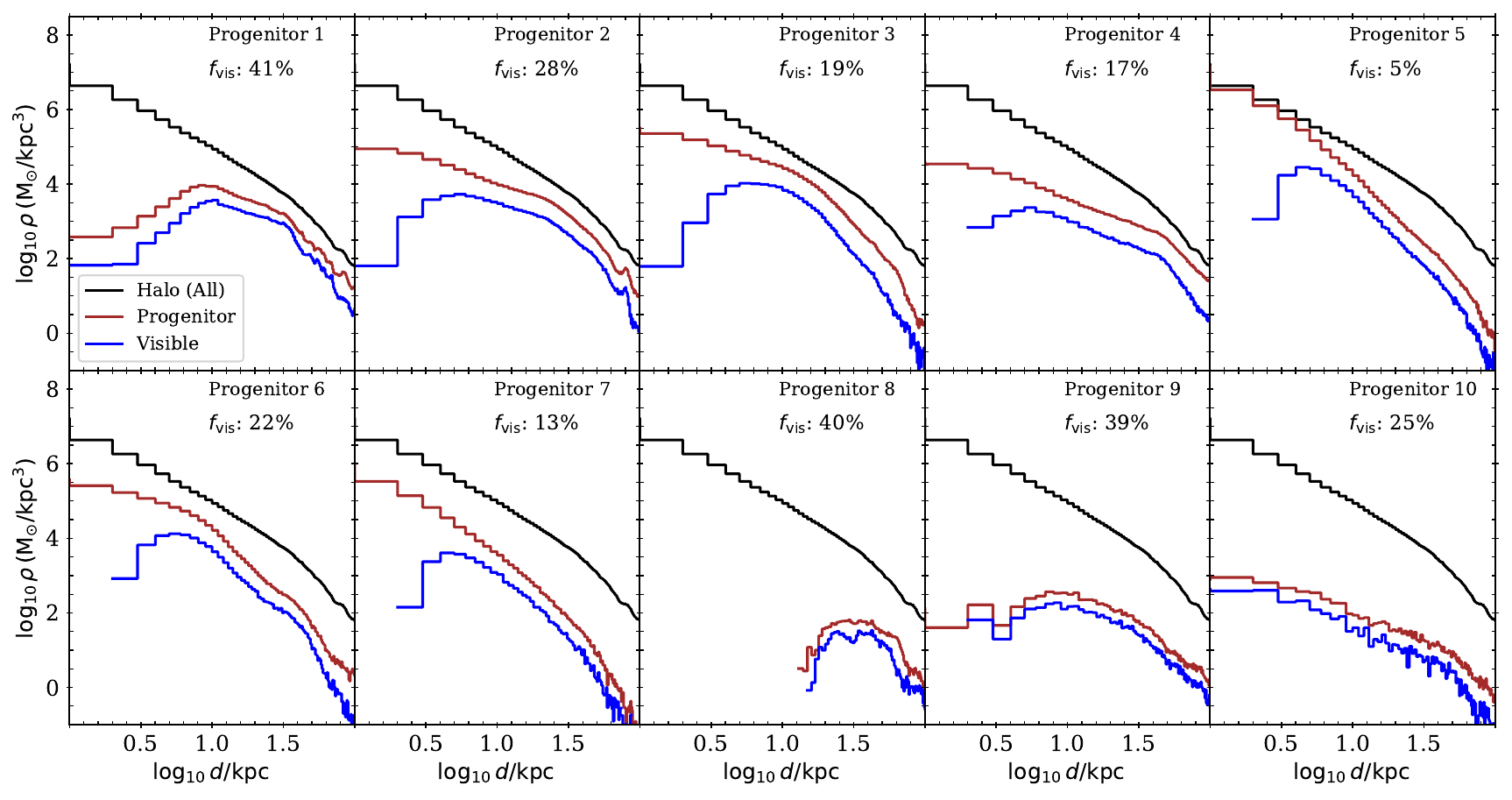}
    \caption{Galactocentric density distribution of star particles originating in the ten most massive accreted progenitors of Au 6. The black line (identical in each panel) shows the density profile of all the accreted star particles. The brown line shows the profile of all star particles from a given progenitor. The blue line shows the profile of star particles that contribute at least one star to the mock catalogue (i.e. to star particles visible within the DESI footprint and selection function). The $f_{vis}$ value quoted in each panel is the fraction of mass visible to DESI.}
    \label{fig:density_tid}
\end{figure*}

Fig.~\ref{fig:density_tid} shows the density profiles of DESI targets associated with the top 10 progenitors, again measured in spherical galactocentric shells, as in Fig.~\ref{fig:density_snap}. The black lines show the total mass density profile for Au 6 (identical in each panel). Each panel corresponds to a different progenitor. The brown lines show the total mass density of all the star particles associated with the progenitor, while the blue lines correspond only to the subset visible to DESI. 

Most progenitors produce centrally concentrated debris extending from the centre of the halo to $\gtrsim100$~kpc. In this example, progenitor 8 stands out as a more concentrated peak in density from $\sim10$ to 80~kpc. This is a coherent tidal stream, reminiscent of the Sagittarius stream in the Milky Way. We discuss this feature in more detail below.  

As in Fig.~\ref{fig:density_snap}, the profiles of total and visible debris in Fig.~\ref{fig:density_tid} are generally similar to each other. This implies that DESI can, in principle, observe an almost unbiased sample of each of the major contributors to the bulk structure of the accreted halo across its full radial extent. In practice, as we discuss below, the finite number of fiber opportunities ($\sim7$ million) that will be allocated to MWS targets means that only a fraction of those visible stars will be observed over the course of the survey. This limits how well MWS can sample very distant features with low surface brightness.

\begin{figure}
    \centering
    \includegraphics[width=\linewidth, clip=True]{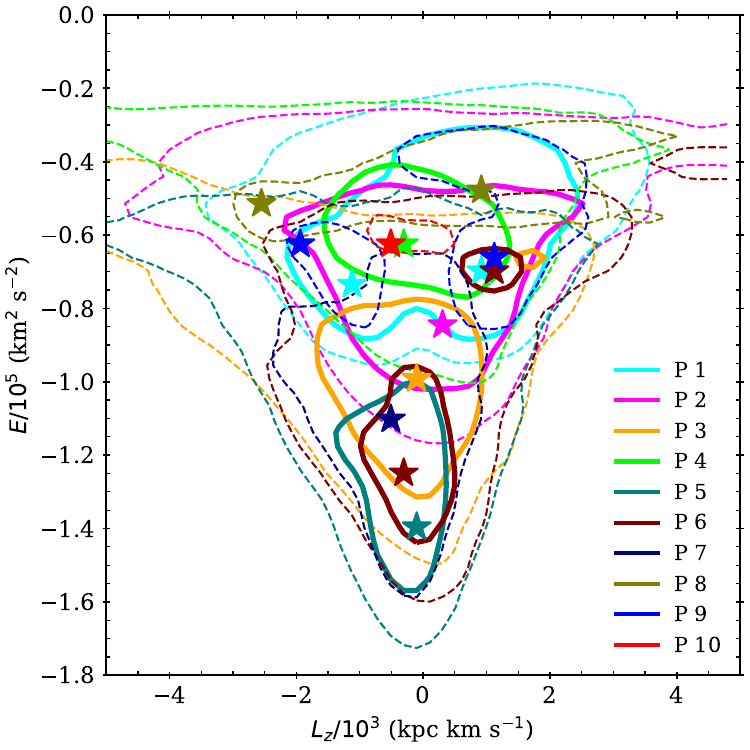}
    \vspace{-4mm}
    \caption{Contours describing the distributions in the space of energy and angular momentum for accreted star particles visible to DESI (see text) associated with each of the top 10 progenitors of Au 6. Dashed and solid contours (in different colours for each progenitor as shown in the legend) indicate regions with a density of 10 particles/bin and 100 particles/bin respectively. With the (arbitrary) choice of bin size in this figure, these correspond approximately to the "outer envelope" and "peak" of each progenitor distribution. Star-shaped symbols of different colours marks the positions of maximum density for each progenitor. Several progenitors (P1, P6, P8 and P9) have two distinct peaks (see also Fig.~\ref{fig:E_Lz_Age_feh}). Prograde rotation (i.e. in the same sense as the galactic disc) corresponds to negative angular momentum. Note that we use Auriga star particles rather than AuriDESI stars to make this figure, as discussed in the text.}
    \label{fig:E_Lz_tid_H6}
\end{figure}

\begin{figure*}
    \centering
    \includegraphics[width=\linewidth, clip=True]{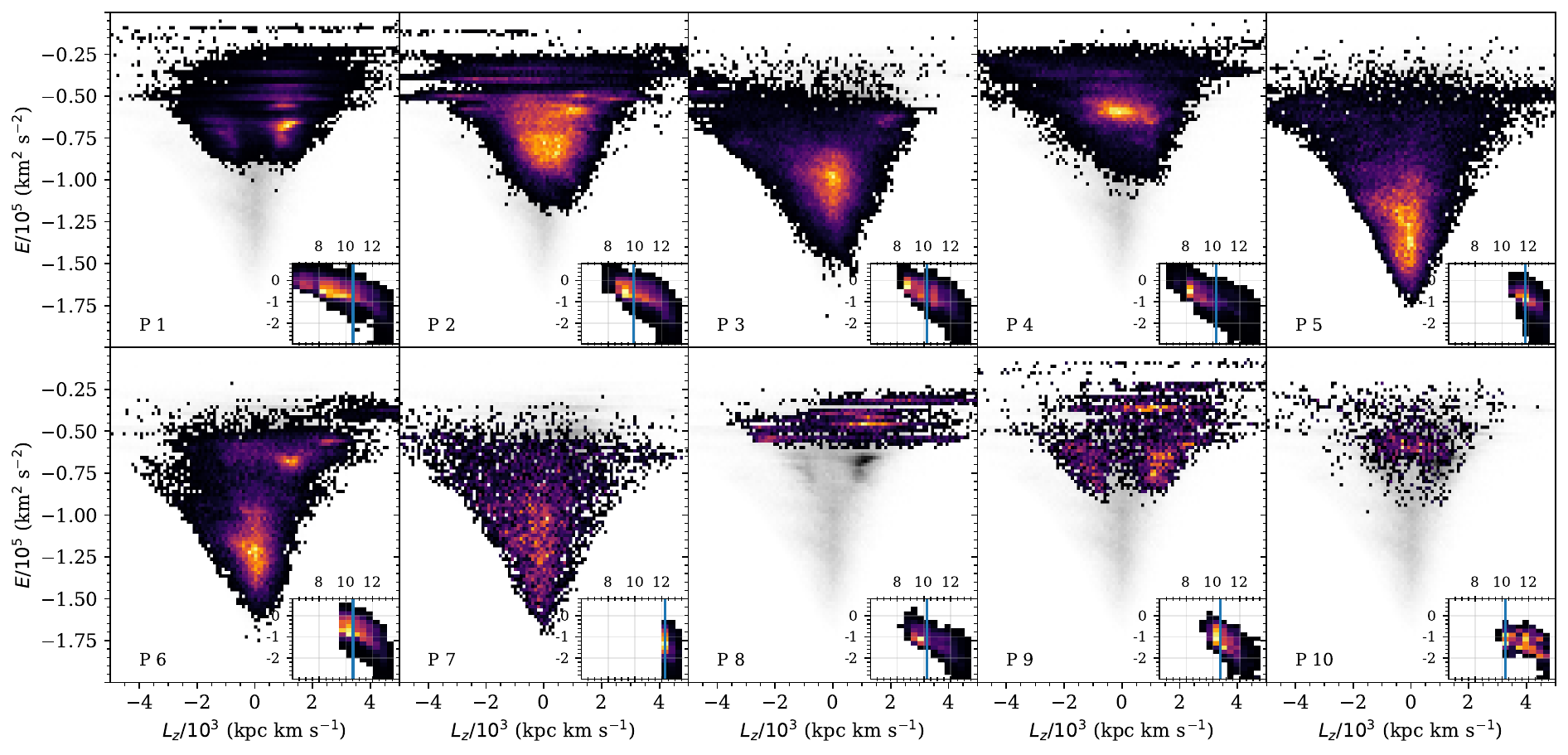}
    \caption{The separate energy -- angular momentum distributions of accreted star particles from the top 10 progenitors. Colours indicate the relative density of stars in each part of the diagram for a given progenitor. For reference, the greyscale distribution in the background of each panel corresponds to the combined density of stars from all progenitors. The inset panels show the distribution of stellar ages (Gyr, horizontal axis) and metallicities (dex, vertical axis). All inset panels have the same range of age and metallicity. The vertical blue line in each inset panel shows the infall time of the progenitor into the main galaxy.}
    \label{fig:E_Lz_Age_feh}
\end{figure*}

Fig.~\ref{fig:E_Lz_tid_H6} shows how the mass accreted from the top 10 most massive progenitors of Au 6, visible given the MWS footprint and selection function, is distributed in the space of total energy and angular momentum (measured about the galactic rotation axis). Contours of individual colours (as shown in legend) enclose the distribution of each of these 10 progenitors, with dashed and solid contours enclosing regions of lesser and higher density of star particles respectively. The region of peak density for each progenitor is marked by a star-shaped symbol with the same colour as the contours of the corresponding progenitor. Few progenitors (P 1, P 6, P 8 and P 9) can be seen with two distinct peaks, either with same energy and opposite signs of angular momentum or at two different energies. Fig. \ref{fig:E_Lz_Age_feh} shows these features in detail. Stars that have been tidally stripped from their progenitors will eventually lose coherence in phase space as they mix along their orbits, but they are expected to preserve approximately their original orbital integrals of motion. Structure in diagrams like Fig.~\ref{fig:E_Lz_tid_H6} has therefore long been considered one of the most promising routes to identifying individual progenitors in the Milky Way's stellar halo \citep{Helmi2000,morrison2009,Carollo2021}. Stars from a given progenitor are also expected to have similar chemical abundances, which may help to distinguish debris with overlapping integrals of motion. However, abundance gradients within the progenitor and accretion spread over multiple pericentric passages may complicate this relationship and pose further difficulties in identifying chemodynamical groups with distinct progenitors \citep{Amarante2022}.

 Most progenitor debris does not show strong rotation; among the rotating debris, there is a weak preference for prograde orbits. As reflected in the relative concentrations of their density profiles (see Fig.~\ref{fig:density_tid}), the progenitors are widely distributed over different regions of the galactic potential well, with progenitor 5 the most strongly bound and progenitors 1, 2, and 4 among the more weakly bound. Except at high energies, there is little apparent structure in this diagram that would be visible without our idealized colour-coding. In reality, only uncertain approximations to progenitor labels (e.g.\ chemical abundances) would be available, and the distribution would also be blurred by observational errors and uncertain knowledge of the Galactic potential. This highlights the fundamental challenge of decomposing the structure of the halo, and the importance of realistic mock catalogues for interpreting surveys on the scale of MWS.

Fig.~\ref{fig:E_Lz_Age_feh} separates the $(E, L_{z})$ diagram into the contribution of each of the top 10 progenitors, and also plots their associated age -- metallicity distributions (inset panels). The blue vertical line in each inset panel shows the infall time, at which the progenitor crossed the virial radius of the central galaxy. These separate diagrams provide a detailed summary of the formation and dynamical history of each progenitor. For example, in the case of progenitor 1, we see two distinct peaks with similar average energies, but opposite signs of angular momentum. Star formation in this progenitor continued long after its infall into the main galaxy, extending from a very early time up to $\sim6$~Gyr before the present, with younger stars being relatively more metal-rich.  Similarly, in the case of progenitor 6, we see a relaxed, deeply embedded, and non-rotating distribution, but also a more weakly bound, retrograde peak with a narrow range of energy and angular momentum. This likely corresponds to the density inflection at $\gtrsim10$~kpc in Fig.~\ref{fig:density_tid}. This progenitor is relatively old overall, with its star formation sharply truncated soon after its infall, at a lookback time of $\sim10$~Gyr. It is nevertheless relatively metal-rich (the bulk of its stars have $\mathrm{[Fe/H]} > -1$).

\begin{figure*}
    \centering
    \includegraphics[width=\linewidth]{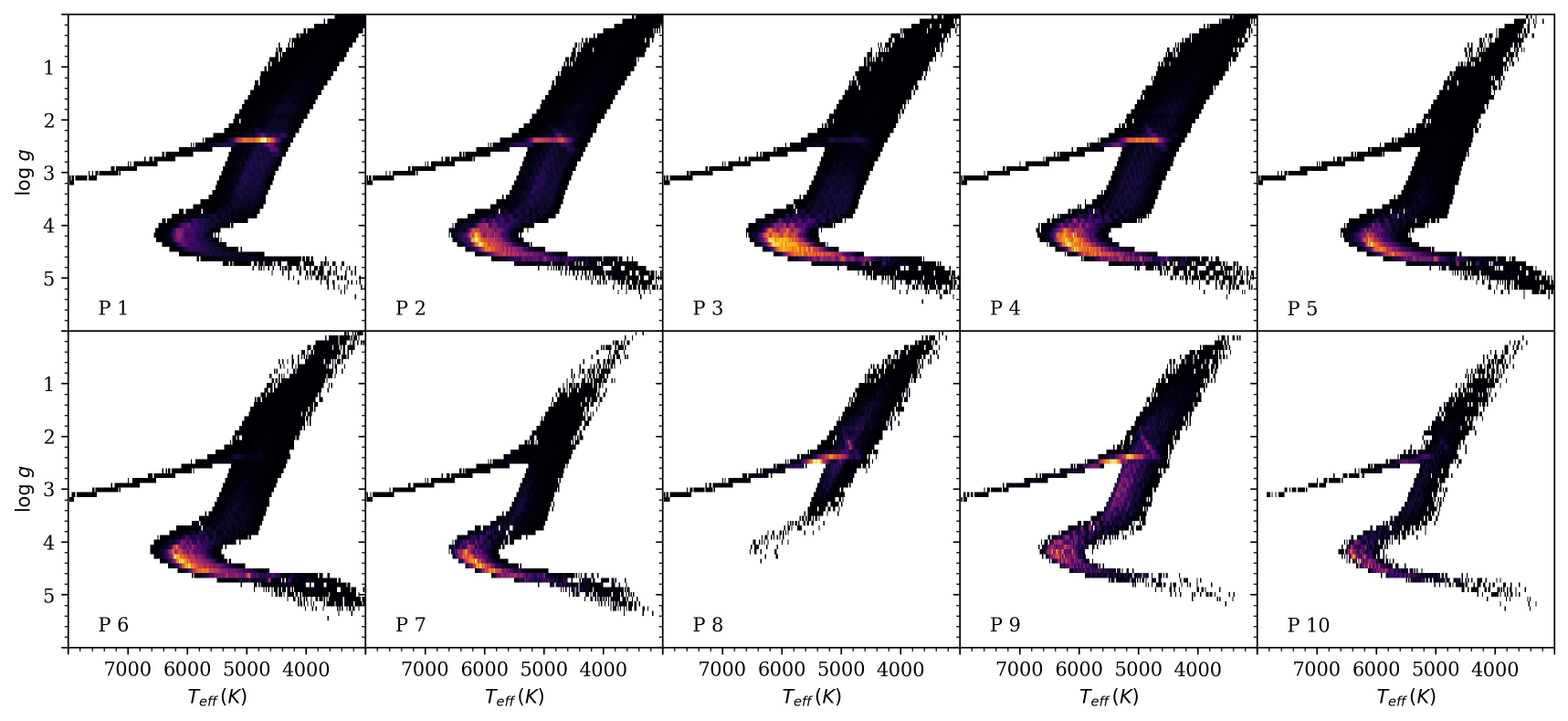}
    \caption{Colour-magnitude distributions of stars in the AuriDESI mock photometric target catalogue for Au 6. The colour scale indicates the relative weight of stars visible to DESI in different evolutionary stages for each progenitor.}
    \label{fig:cmd_tid}
\end{figure*}

Fig.~\ref{fig:cmd_tid} shows the colour-magnitude diagram of the individual stars from each progenitor in the AuriDESI photometric target catalogue. Different evolutionary stages enter the DESI target selection at different distances.
Most of the stars observable by DESI are close to the main sequence and turn-off (MSTO). The full CMD is accessible for all but progenitor 8, as expected given the wide radial range of debris in each case (shown in Fig.~\ref{fig:density_tid}). Progenitors 1, 2, 4, and 9 appear with a clear red clump and MSTO; other progenitors are represented mainly by their MSTO. For progenitor 8, which covers a much smaller range of distance, stars fainter than the sub-giant branch fall below the MWS magnitude limit. The CMDs are noticeably narrower for the less massive progenitors, which have a narrower range of ages and metallicities. 

\begin{figure*}
    \centering
    \includegraphics[width=15cm]{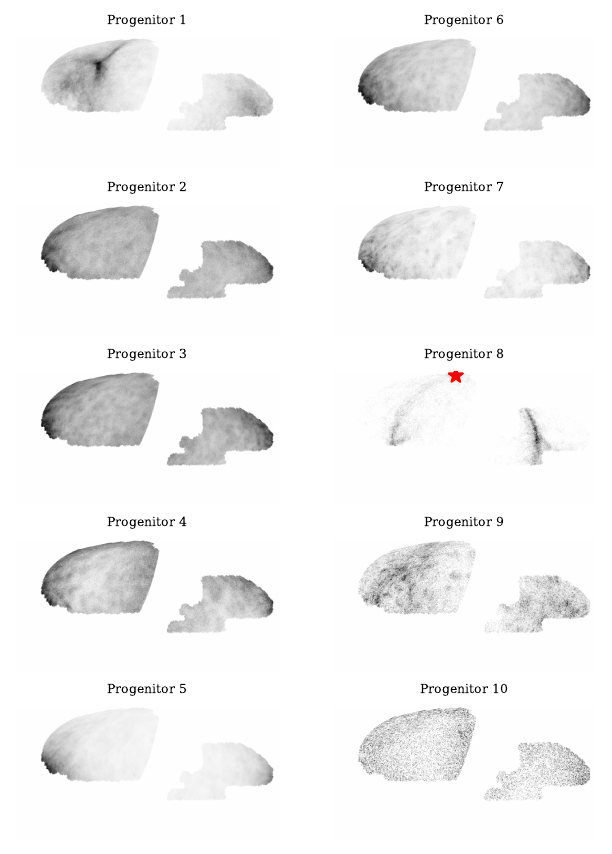}
    \caption{Distribution of stars from the 10 most massive progenitors in equatorial coordinates in the Au 6 AuriDESI mock catalogue. Clear stellar substructures are visible for some progenitors. The red star marked on Progenitor 8 represents the position of its bound satellite remnant (see section \ref{Satellites}).}
    \label{fig:skyplot_tid}
\end{figure*}

Finally, Fig.~\ref{fig:skyplot_tid} shows sky projections of the debris for the top 10 progenitors. In most cases, the debris is smoothly distributed. The most obvious exception is the debris from progenitor 8. This comprises a coherent stream on a polar orbit, somewhat resembling the Sagittarius stream as noted above (Fig.~\ref{fig:density_tid}, \ref{fig:cmd_tid}). Like Sagittarius, the core of this progenitor is still intact, marked by a red star symbol in Fig.~\ref{fig:skyplot_tid}. Progenitor 1 also shows a coherent overdensity in configuration space, although (according to SUBFIND) this is not a bound remnant of the progenitor.

To illustrate halo-to-halo variations, Appendix \ref{halo21_appendix} repeats some of the plots in this section for Au 21.

\section{Applications of the AuriDESI mocks}
\label{sec:applications}

\begin{figure*}
    \centering
    \includegraphics[width=\linewidth]{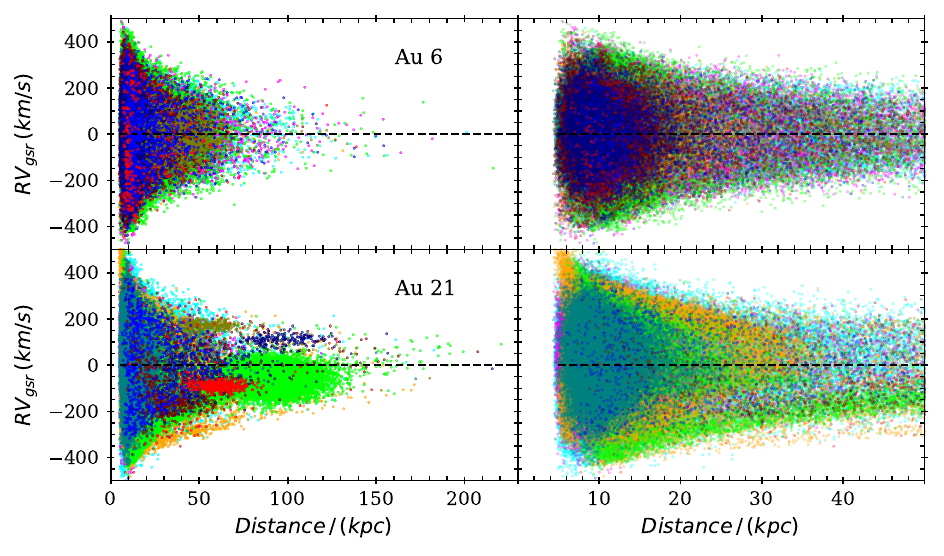}
    \caption{Radial velocity distribution of stars in AuriDESI Au 6 (top) and Au 21 (bottom) with galactocentric distance. The left panel shows the distribution for all the stars in the halo while the right panel shows the distribution for stars within a distance of 50 kpc. Stars in the top 10 most massive progenitors are plotted with markers of different colours (the colour scheme is the same as previous figures). The black dashed line indicates 0 $\mathrm{km\,s^{-1}}$ radial velocity. The dense structures visible in Au 21 (left panel) are satellite remnants. The distances of stars have been convolved with a fiducial Gaussian error of 15\%.}
    \label{fig:vrad_dist_FA}
\end{figure*}

\subsection{Spectroscopic forecasts for the complete Milky Way Survey} 
\label{forecasts}
So far, we have discussed the scope of mock DESI observations for one Auriga galaxy, Au 6, based for the most part on the AuriDESI mock photometric target catalogue. We now consider the mock spectroscopic data. As described in Section \ref{spectroscopic catalogue}, stars are selected for mock-spectroscopic observation by applying the DESI fiber assignment algorithm to the mock photometric target catalogue. The result is a mock dataset corresponding to the stars that will be observed during the 5-year DESI main survey, for which DESI will obtain radial velocities, atmospheric parameters, and metallicities. This section forecasts, in broad terms, some of the results that may be obtained from the full MWS dataset, based on these mock observations of the Auriga galaxies.

Fig. \ref{fig:vrad_dist_FA} shows the distribution of stars in the space of radial velocity and galactocentric distance, colour-coded by the progenitor (among the 10 most massive) from which the star was  accreted. We convolve all distances with a fiducial 15\% Gaussian error\footnote{In practice, distances for DESI stars will be estimated using a variety of spectrophotometric and astrometric methods, the accuracy of which will depend on the type of star as well as the distance. Such distances will be a value-added product of the DESI survey, rather than a primary spectroscopic measurement.}. The left panel shows the whole distribution, while the right panel shows a zoomed-in version for stars at distances less than 50 kpc. This plot reveals different kinematic structures within the footprint of DESI. In Au 6, most of the progenitors are well-mixed in this space, whereas in Au 21, regions dominated by individual progenitors are apparent. These are due to satellites that are still in the process of being tidally disrupted. The third most massive progenitor of Au 21 (in orange, as in previous figures) has multiple orbital wraps (see right panel) and an infalling tail extending to large distances. We discuss the properties of this progenitor further in appendix \ref{halo21_appendix}.

Fig. \ref{fig:met_dist_fa} shows the metallicity profile of stars in bins of Galactocentric radius. This shows the differences between true and observed metallicity profiles. The solid lines represent the metallicity distribution of observed stars while the dashed lines show the true metallicity of star particles. Again, we have convolved the distances with a fiducial Gaussian error of 15\%. The peaks in the profiles of Au 16, Au 21, and Au 27 correspond to satellite remnants. At all distances, the true mean metallicity of the star particles is higher than that measured by our mock surveys (except for Au 21, at the position of its massive satellite remnant). 

To understand how this effect arises from the MWS selection function, we consider one example in more detail. Fig. \ref{fig:met_dist_fa_target} is an extension of Fig.\ \ref{fig:met_dist_fa}, focusing on Au 6. The black dotted line represents the metallicity profile of all the star particles, while the black dashed line represents that of all the stars in Au 6. The stars are separated into \mainblue{} and \mainred{} (the spectroscopic catalogue contains very few \mainbroad{} stars, see section \ref{discussion} for details), and the \mainred{} targets are further subdivided based on their colour (as indicated by red solid and dashed lines). Although the true and observed metallicity profiles are very similar up to $\sim 75$ kpc, the difference is much higher at larger distances. At smaller distances, the thin disc comprises a relatively metal-rich, young stellar population. As seen in Fig. \ref{fig:dist}, due to the relatively lower number of thin disc stars in AuriDESI, stars selected by the \mainblue{} sample dominate the mock spectroscopic survey up to $\sim 10$ kpc. In the $\sim 10 $ -- $30$ kpc region, both \mainblue{} and \mainred{} stars make similar contributions, while at larger distances \mainred{} stars become dominant. The \mainblue{} target class primarily consists of metal-poor MSTO stars in the thick disc and stellar halo (including metal-poor BHB stars up to $\sim$ 50 kpc), leading to an average metallicity for observed stars that is lower than the underlying average for star particles at those distances. At larger distances, the \mainred{} targets that dominate the sample are predominantly 
metal-poor giants. From the figure, we see that, at larger distances, the observed metallicity profile follows the profile of \mainred{} stars. The significant difference at distances greater than 100 kpc can therefore be attributed to the magnitude limits imposed by the MWS selection function; metal-rich red giants at these distances are typically fainter than the magnitude limit of DESI. 

Fig. \ref{fig:E_Lz_FA} shows the distribution of Au 6 stars from the mock spectroscopic survey in energy-angular momentum space. In the left panel, stars from the 10 most massive progenitors are plotted individually, with different colours. In the right panel, contours indicate the distribution of stars from each progenitor. Every progenitor possesses its own characteristic orbital integrals of motion, which are approximately conserved by its stars as they mix with those from other progenitors in configuration and velocity space. In principle, this diagram therefore provides important insights into the assembly history of a galaxy \citep{Helmi2000,Grand2019,Christine2019}. However, Fig. \ref{fig:E_Lz_FA} highlights the difficulty of distinguishing individual progenitors in this space in practice. From both the panels it can be seen that multiple progenitors occupy any given region of the E-Lz space. In the mock catalogues, we can "colour code" stars according to their known progenitor; in real life, this coding can only be approximated, for example with labels derived from kinematics and chemical abundances. The different versions of this diagram throughout this paper (Fig. \ref{fig:E_Lz_tid_H6},\ref{fig:E_Lz_Age_feh},\ref{fig:E_Lz_Prog}) illustrate 
 both the scientific potential of a large homogeneous survey like MWS and the considerable challenge associated with the recovery of individual progenitors. It is clear from these figures that a survey like MWS can yield a dense sample of almost all the most significant accretion events that have contributed to the stellar halo, although much remains to be understood about how to extract and interpret this information. The AuriDESI mock catalogues provide a realistic basis for developing such analyses in the context of DESI.

Finally, we make a simple estimate of the variance in a DESI-like survey due to the observer's azimuthal position in the Galaxy. Fig.\ \ref{fig:bhb_fa} shows the distribution of BHBs (as defined in section \ref{spectroscopic catalogue}) in the six AuriDESI spectroscopic mock catalogues, beyond a  heliocentric distance of 10 kpc. The different line styles represent the four different angular positions of the Sun within the galactic disc for which we provide mock realizations ($30^{\circ}, 120^{\circ}, 210^{\circ}$, and $300^{\circ}$). From the figure, it is clear that, although there are minor differences between the different observer locations, most obviously at very large distances, the overall distribution is almost identical. BHBs are arguably one of the most useful tracers of the bulk structure of the halo; although they are sparse, their distances can be measured accurately and they are selected by MWS over a broad range of distance. From this figure we conclude that, at least in the mock catalogues, MWS samples a wide enough area to ensure that the overall spatial and kinematic structure of the metal-poor halo can be measured robustly.
\begin{figure}
    \centering
    \includegraphics[width=\linewidth]{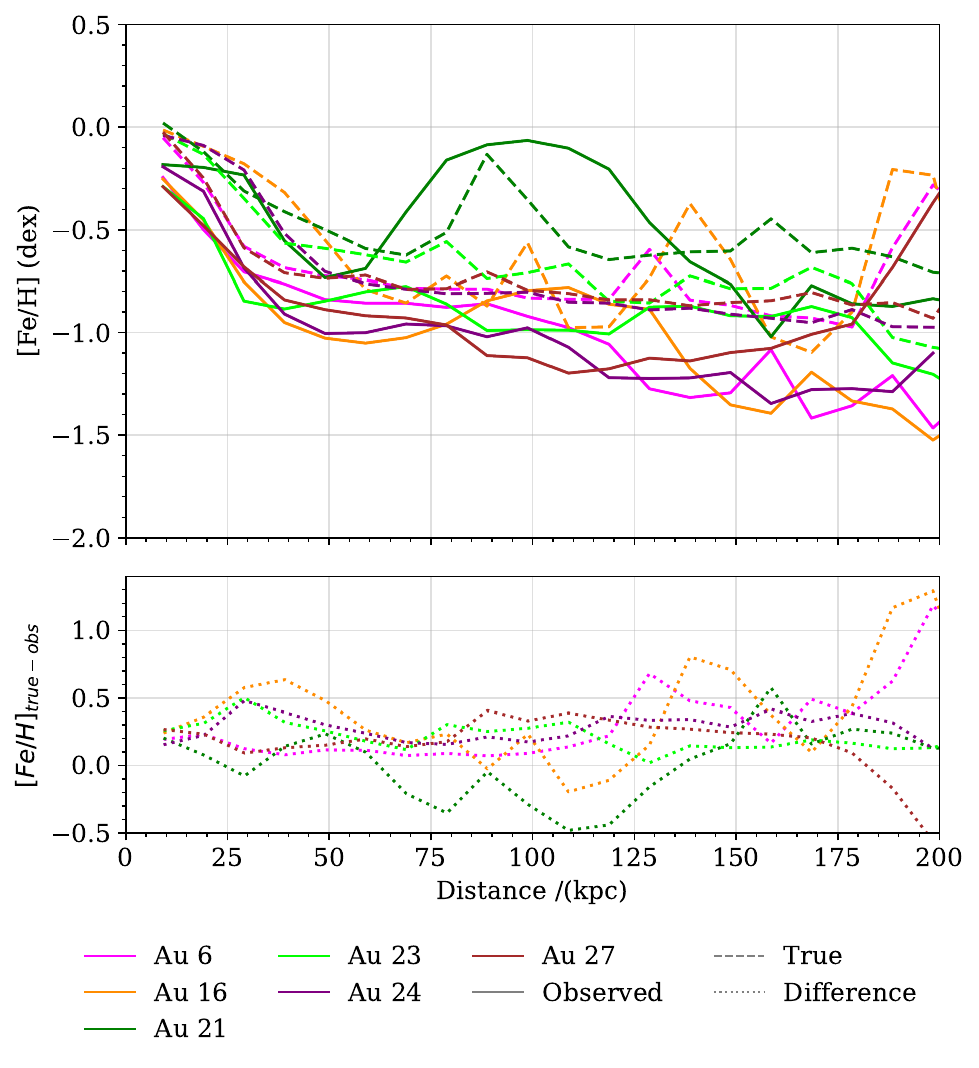}
    \caption{Metallicity in bins of galactocentric distance. Solid lines represent the metallicity profile of stars (referred to as "observed") while dashed lines represent that of all simulation star particles (referred to as "true"). The distances of stars have been convolved with a fiducial Gaussian error of 15\%.}
    \label{fig:met_dist_fa}
\end{figure}

\begin{figure}
    \centering
    \includegraphics[width=\linewidth]{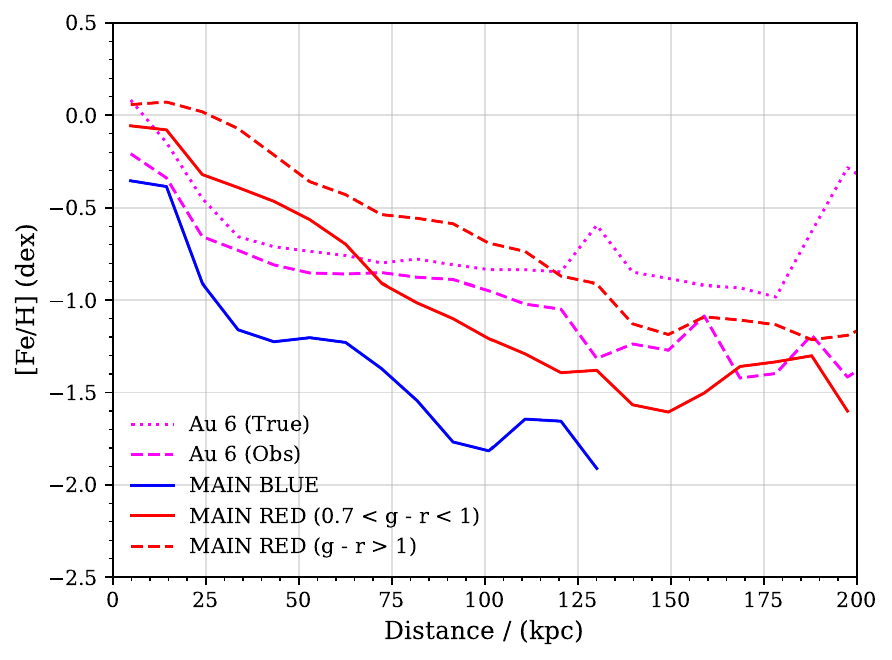}
    \caption{Metallicity in bins of galactocentric distance for AuriDESI Au 6. The black dotted line represents all the star particles (referred to as "true") in Au 6 (same as the blue dashed line in Fig.\ \ref{fig:met_dist_fa}) and the black dashed line represents stars (referred to as "Obs") in the spectroscopic catalogue of AuriDESI (same as the blue solid line in Fig.\ \ref{fig:met_dist_fa}). Blue and red (dashed and solid) lines represent the metallicity distribution of \mainblue{} and \mainred{} target classes. The distances of stars have been convolved with a fiducial Gaussian error of 15\%.}
    \label{fig:met_dist_fa_target}
\end{figure}

\begin{figure}
    \centering
    \includegraphics[width=\linewidth]{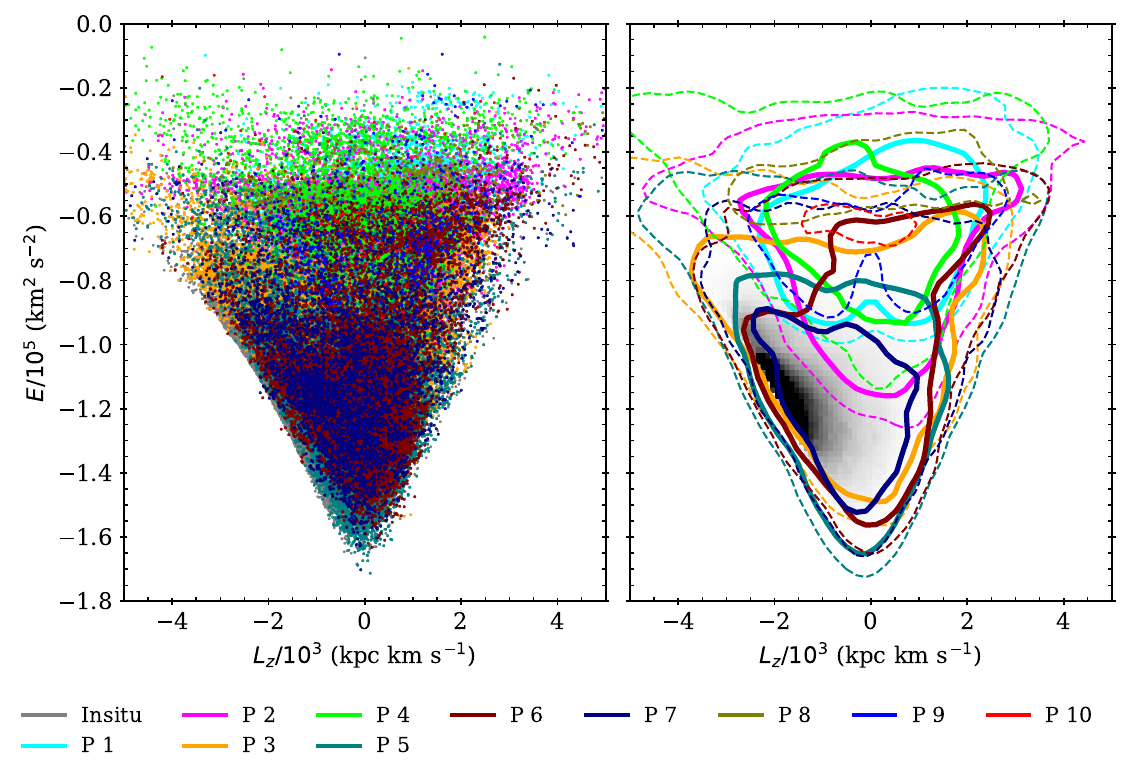}
    \vspace{-4mm}
    \caption{Two alternative views of the energy -- angular momentum distribution of stars in the Au6 AuriDESI spectroscopic catalogue. Left panel: Stars from the top 10 most massive progenitors are plotted as individual points with different colours. Right panel: Contours describing the distribution of stars shown in the left panel. Dashed (and solid) contours enclose regions of lesser (and higher) density (10 and 100 stars per bin respectively) associated with each progenitor.  In both panels, the distribution of \insitu{} stars is shown in the background, with the majority distributed along the locus of prograde near-circular orbits in the galactic disc. }
    \label{fig:E_Lz_FA}
\end{figure}

\begin{figure}
    \centering
    \includegraphics[width=\linewidth]{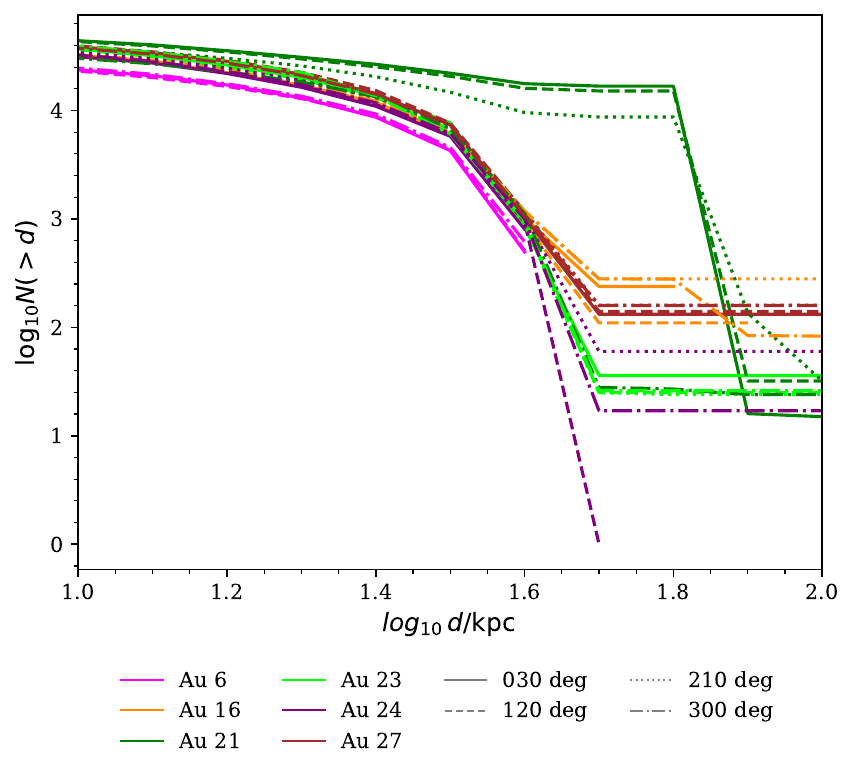}
    \caption{Distance distribution of BHB stars selected from the AuriDESI spectroscopic mock catalogues. The different line styles represent different angular positions of the Sun within the galactic disc (30$^{\circ}, 120^{\circ}, 210^{\circ}, 300^{\circ}$).}
    \label{fig:bhb_fa}
\end{figure}

\subsection{Comparison with DESI SV3 data}
\label{sec:sv3}

In this section, we compare our AuriDESI Au 6 mock to real DESI data for stars observed in SV3, a program of  observations over $\sim100$~sq.\ deg.\ carried out prior to the start of the main DESI survey in order to validate the target selection, survey design and data quality (see Section \ref{spectroscopic catalogue}). Real data taken in SV3 and reduced with the MWS pipeline are available in the DESI early data release \citep{EDRvalidation2023,EDRadame2023}.

The observing strategy for SV3 was quite different to that used in the main survey. SV3 employed a large number of overlapping DESI tiles to densely sample small patches of sky. It would therefore not be appropriate to construct a mock SV3 sample simply by drawing targets from an AuriDESI mock spectroscopic catalog in the areas covered by SV3. The correct approach would be to pass the AuriDESI photometric target catalog to the DESI fiber assignment algorithm again, for the specific set of tiles used for SV3, rather than the tiles used in the main survey. However, the SV3 fields have $\gtrsim90\%$ spectroscopic completeness \citep[see][]{2022APC}. For simplicity, we therefore approximate the results of fiber assignment in SV3 by randomly sampling AuriDESI targets in the SV3 areas to match the total number of DESI SV3 stars observed in each of the \mainblue{}, \mainred{} and \mainbroad{} target categories\footnote{Since the Auriga haloes are more massive than the real MW, the number of stars in AuriDESI within the SV3 fields is an order of magnitude higher than that in the real observations; were the DESI SV3 observations carried out in the AuriDESI haloes, they would have much lower completeness for the same fiber budget. This is why we randomly sample to the same density as the real observations, rather than taking (say) $90\%$ of all the AuriDESI targets in the SV3 areas.}.

\begin{figure}
    \centering
    \includegraphics[width=\linewidth]{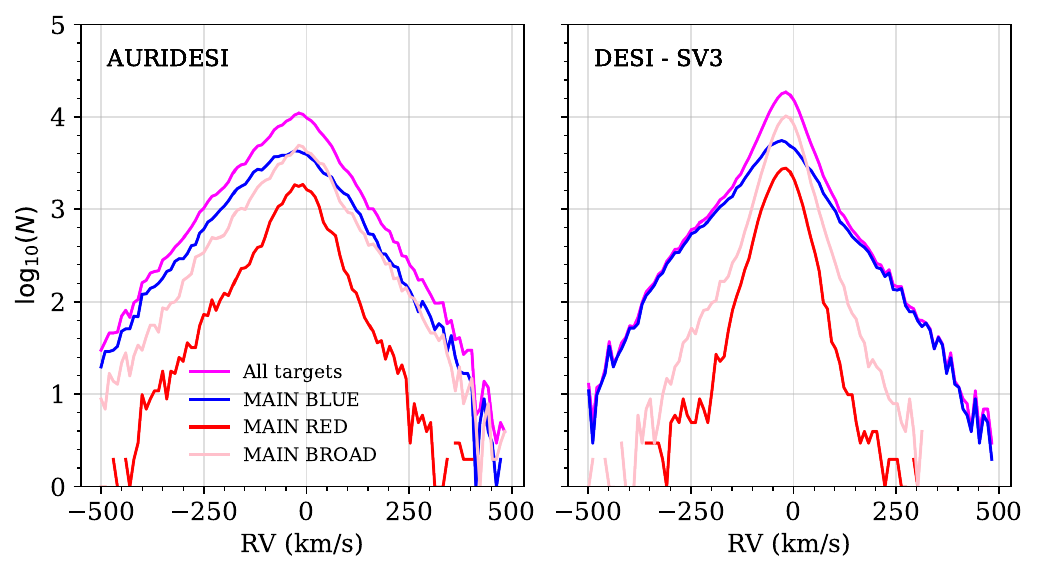}
    \caption{The radial velocity distribution of AuriDESI Au 6 stars within the DESI SV3 footprint (left panel) compared to that of stars observed by DESI during its SV3 program (right panel). Stars are divided into \mwsmain{} target classes (see Appendix~\ref{name:mws selection criteria}). For each target class, the AuriDESI data have been randomly sampled to obtain the same total count as the SV3 data, as described in the text.}
    \label{fig:vrad_desi_auri}
\end{figure}

\begin{figure}
    \centering
    \includegraphics[width=\linewidth]{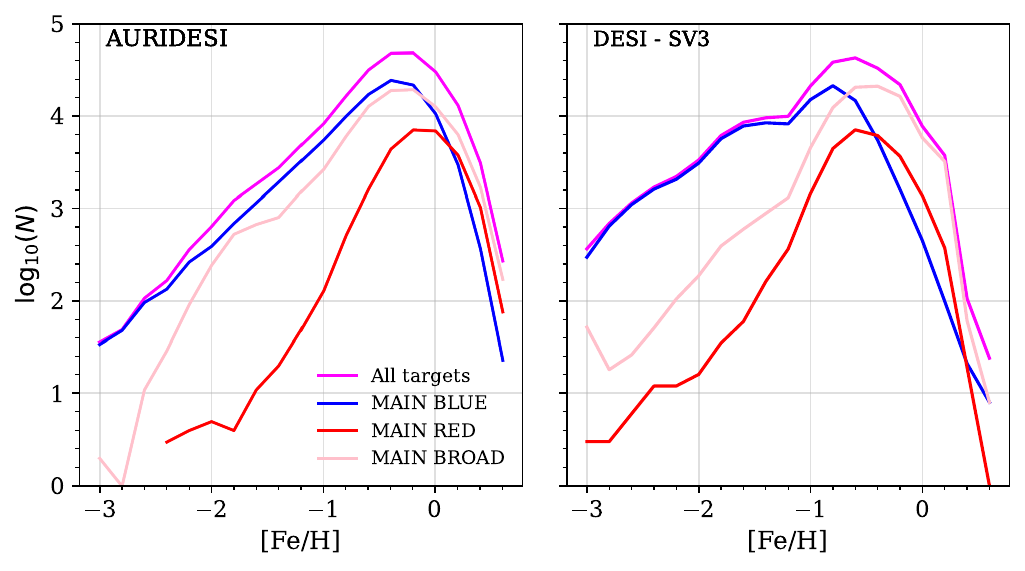}
    \caption{The metallicity distribution of AuriDESI Au 6 stars  within the DESI SV3 footprint (left panel) compared to that of stars observed by DESI during its SV3 program (right panel). Stars are divided into \mwsmain{} target classes (see Appendix~\ref{name:mws selection criteria}). The AuriDESI stellar haloes are more metal-rich than the stellar halo of the real Milky Way. The mock data have been down-sampled to match the SV3 counts for each target class, as in Fig.~\ref{fig:vrad_desi_auri}.}
    \label{fig:feh_desi_auri}
\end{figure}

Fig. \ref{fig:vrad_desi_auri} shows the radial velocity distributions of different MWS target classes, comparing AuriDESI Au 6 to MWS EDR data within the SV3 footprint. The distributions for AuriDESI look similar to those in the real Milky Way overall. However, their characteristic widths differ: the standard deviations of the distributions from AuriDESI are $(117, 79, 103)\,\mathrm{km\,s^{-1}}$, for \mainblue{}, \mainred{} and \mainbroad{} respectively, whereas the corresponding values for SV3 are  $(157, 55, 49)\,\mathrm{km\,s^{-1}}$. This may be due to a difference between the potential of the Au6 system and that of the Milky Way, although the virial mass of Au6 is close to typical estimates for the Milky Way (see Table~\ref{tab:table1}). Alternatively, they may arise from differences in the star formation and tidal mass loss histories of individual satellites, both of which affect the number of stars on orbits with high radial velocity. The number of \mainbroad{} stars in the AuriDESI SV3 fields is slightly lower than that observed by DESI, despite our attempt to match the number in the real SV3 dataset by construction, due to the lack of low-mass stars in the thin disc of the Auriga simulation (as shown in previous figures, see appendix \ref{isochrones} for more details).

Fig. \ref{fig:feh_desi_auri} shows the metallicity distribution for different MWS target classes, again comparing AuriDESI Au 6 to MWS data within the SV3 footprint. This plot clearly shows that AuriDESI is more metal-rich than the real Milky Way. \mainblue{} stars, which include the metal-poor halo, has a peak at $\approx -0.2$~dex for AuriDESI and $\approx -0.8$~dex  for observed SV3 stars. \mainred{}, which mainly targets the thick disc giants and turn-off stars, has a peak at $\approx -0.1$~dex  for AuriDESI and $\approx -0.6$~dex  for SV3. \mainbroad{}, which targets the metal-rich thin disc, has a peak at $\approx -0.2$~dex for the mock and $\approx -0.5$~dex for SV3.  

\begin{figure}
    \centering
    \includegraphics[width=\linewidth]{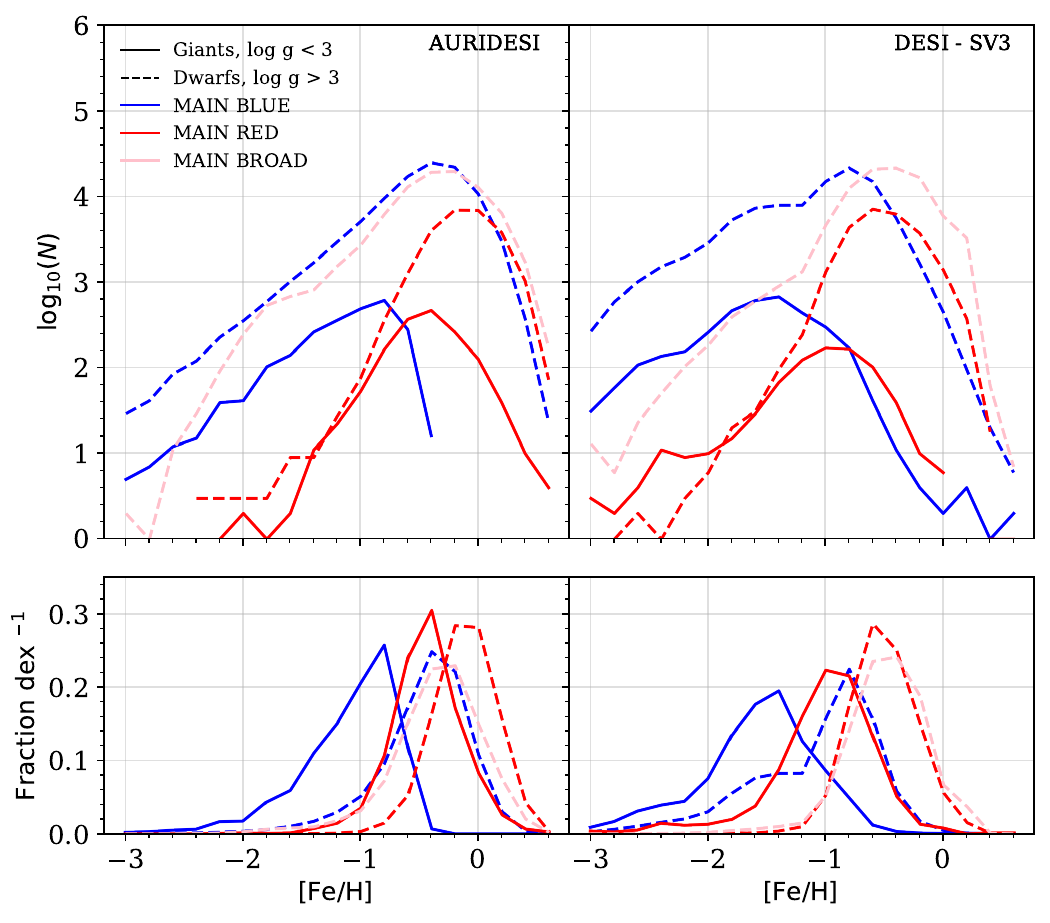}
    \caption{A comparison of the metallicity distributions for  stars in AuriDESI Au 6 within the SV3 footprint (left) and real stars from DESI SV3 program (right). Stars are divided into different target classes and further separated into giants (solid lines) and dwarfs (dashed lines). The top panels show absolute numbers of stars per metallicity bin on a logarithmic scale; the bottom panels show the corresponding fractions of stars in a given category per bin, on a linear scale.
    The mock data have been down-sampled to match the SV3 counts for each target class, as in Fig.~\ref{fig:vrad_desi_auri}.} 
    \label{fig:feh_giants_dwarfs}
\end{figure}

The top panel of Fig. \ref{fig:feh_giants_dwarfs} again shows the metallicity distribution of stars in different target classes, now separated into giants and dwarfs. The bottom panel shows the relative distribution of metallicities, showing the percentage of samples in different categories with metallicities in bins of 0.2 dex. Separating the metallicity distributions of dwarfs and giants provides a consistency check on the $\log\,g$ measurements: we expect giants in the MWS selection to be mostly in the stellar halo, and hence more metal poor than dwarfs in the same selection, and we expect redder giants to be more metal rich than bluer giants. These differences should also be apparent in the mock data. Fig.~\ref{fig:feh_giants_dwarfs} shows that \mainblue{} giants have a median [Fe/H] of $\approx -1.5$ (observed) and $\approx -0.73$ (mock). \mainred{} giants are metal-rich compared to \mainblue{} and have a median [Fe/H] of $-0.89$ (observed) and $-0.21$ (mock). All three target classes show similar offsets between the metallicity distributions of AuriDESI and real DESI data, apparent in figures \ref{fig:feh_desi_auri} and \ref{fig:feh_giants_dwarfs}. 

\begin{figure}
    \centering
    \includegraphics[width=\linewidth]{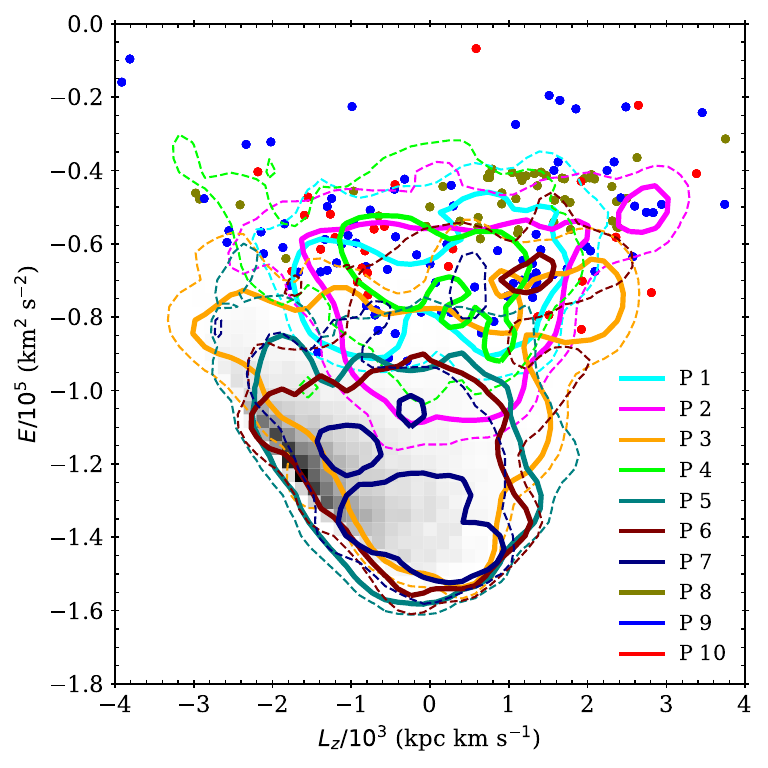}
    \vspace{-4mm}
    \caption{Contours describing the distribution in the energy and angular momentum space for stars associated with each of the top 10 most massive progenitors in Au 6, within the SV3 fields. Dashed (and solid) contours encloses regions of lesser (and higher) density of stars (10 and 100 stars per bin respectively) associated with each progenitor (marked in different colours as shown in the legend). Stars from less massive progenitors (P 8, P 9 and P 10) are shown with individual points. The distribution of \insitu{} stars within the SV3 fields is shown in the background; the majority of the \insitu{} mass is in the (prograde) galactic disc, which traces the minimum envelope of energy for negative angular momenta. In this figure, stars from the mock have been randomly sampled in order to reproduce the observed SV3 counts in each target class, as described in the text.}
    \label{fig:E_Lz_Prog}
\end{figure}
Fig. \ref{fig:E_Lz_Prog} shows the distribution in energy-angular momentum space for stars in the SV3 footprint, restricted to those stars belonging to the 10 most massive progenitors in Au 6. For massive progenitors, we use contours (with individual colours shown in the legend) to mark the envelope of their distribution and the region(s) of their highest density in this space. For the three least massive of these progenitors (P8, P9 and P10) we use individual points. The distribution of the \insitu{} population is shown in the background. Although the \insitu{} population has a similar distribution to the accreted stars overall, the majority of the \insitu{} stars are concentrated along on the left edge of the distribution (i.e. on approximately circular orbits, in the galactic disc). This figure illustrates the substantial amount of information available within the relatively small footprint of SV3 ($\sim100$ square degrees in total, distributed over the MWS footprint). In this halo, no single progenitor dominates this distribution; on the contrary, even an SV3-like survey samples many of the major substructures in the halo. The differences seen in the structure of this plot, when compared to figure \ref{fig:E_Lz_tid_H6}, could be the result of the placement of individual SV3 fields.

\subsection{Satellites in AuriDESI} 
\label{Satellites}

The morphology and kinematics of tidal streams and other coherent features that arise from satellite accretion events are sensitive to the gravitational potentials of the primary galaxy and the infall time and orbit of the progenitor \citep[e.g.][]{Amorisco2017}. In some cases, progenitor satellites may survive to the present day despite significant tidal stripping. The orbits and internal structure of these remnant progenitors may be different from those at infall, and they may not show obvious signs of ongoing mass loss \citep{MYWang2017,Shipp2023}.

The AuriDESI mock catalogues contain stars associated with progenitor satellites that survive to the present day. As mentioned in Section~\ref{sec:satellite_labels}, bound satellites can be identified with a unique non-zero values of the \texttt{SubhaloNr} column in the catalogue. All stars that originate in a given satellite will, by construction, have the same \texttt{TreeID} value. Stars that have been tidally stripped from a surviving satellite will have therefore have the same \texttt{TreeID}, and can be identified as bound to the main halo rather than their parent satellite by having \texttt{SubhaloNr = 0}. Since the mock catalogue only contains stars targeted by DESI, it is possible that tidally stripped stars associated with a surviving satellite may appear in the catalogue even though their parent satellite does not (this may happen, for example, if the satellite falls outside the DESI footprint). For a similar reason, the information available in the mock alone is not sufficient to compute (for example) the total stellar mass of a progenitor reliably, regardless of whether it survives as a satellite or not (only the brightest stars will be included in the mock). We therefore provide a table of properties for all the progenitors of each halo, surviving and disrupted (described in Appendix \ref{tab:Columns in PROGENITOR extension in AuriDESI}). 

\begin{figure}
    \centering
    \includegraphics[width=\linewidth]{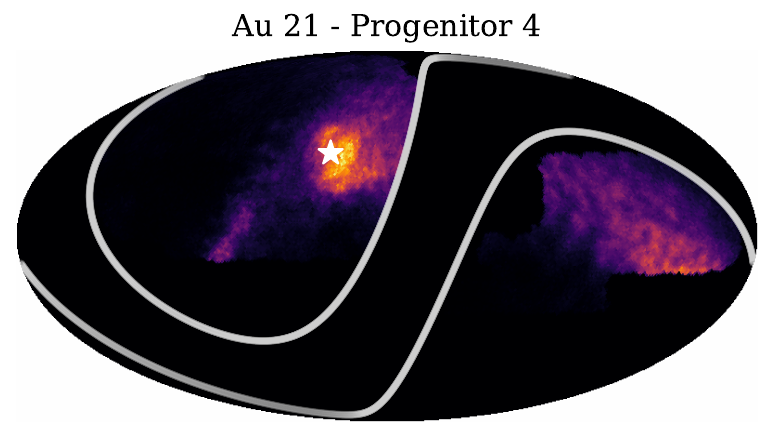}
    \caption{Density of stars originating from progenitor 4 of Au 21 in equatorial coordinates. The white star symbol marks the position of the surviving satellite. The white lines represent the galactic latitude limits of the DESI survey at |b| = $\pm 20^{\circ}$. }
    \label{fig:Skyplot_P4_H21}
\end{figure}

In section~\ref{Halo 6} we described the top 10 progenitors in Au 6, including one example of a tidal stream associated with a surviving satellite (progenitor 8, marked with a red star symbol in Fig.~\ref{fig:skyplot_tid}). This progenitor has a total stellar mass of $3.6\times 10^7\,\msol$, of which $3.16 \times 10^6\,\msol$ is visible to DESI (according to the definition used in Section~\ref{sec:top_10_composition}) at the present day. Fig.~\ref{fig:Skyplot_P4_H21} shows another example, progenitor 4 of Au 21 (similar figures for all the top 10 progenitors in this halo are given in Appendix \ref{fig:H21_Progenitors}). The figure shows a very prominent region of high overdensity in the DESI footprint, corresponding to the satellite body (again marked with a star symbol) and a broad stream extending across the footprint. This progenitor has a total (star particle) mass of $5.8 \times 10^{9}\,\msol$. The bound  satellite remnant retains approximately 84  per cent of this total, i.e. $4.9 \times 10^{9}\,\msol$, of which $2.7 \times 10^{9}\,\msol$ is visible to DESI. In Au 21, 6 out of the 10 most massive progenitors of the accreted stellar halo have surviving satellite remnants within the DESI footprint, some with masses as low as $6.06 \times 10^{4}\, \msol$ (progenitor 5 in figure \ref{fig:H21_Progenitors}). 

Results such as these, albeit only for a small sample of simulated halos, suggest that we might expect more than one prominent overdensity in the Milky Way halo to be associated with a surviving progenitor galaxy, in addition to the well-known example of the Sagittarius stream. Similar results are reported by \citet{MYWang2017}, based on the APOSTLE simulations, and by \citet{Shipp2023}, based on mock Dark Energy Survey (DES) observations using the FIRE simulations. Together with the supplementary information we describe in described in Appendix \ref{tab:Columns in PROGENITOR extension in AuriDESI}, the AuriDESI mock catalogues can be used to explore how DESI MWS can link debris structures to surviving satellites and constrain the infall times and masses of disrupted progenitors.

\section{Discussion and Conclusion}
\label{discussion}

We have presented AuriDESI, a set of mock catalogues that match the observation footprint and selection functions of the DESI Milky Way Survey. These are based on the Auriga cosmological simulations of Milky Way analogues and build on the AuriGaia mocks catalogues of \citet{Grand2018}. 
We provide AuriDESI mock catalogues for 6 haloes and 4 alternative solar positions in each halo. The data model of the mocks follows closely that of the real DESI data products (Koposov et al. (in preparation)): photometric target catalogues and tables of spectroscopic pipeline measurements for the subset of targets that will be observed in the 5-year DESI bright time survey. The six mock catalogues presented in this paper will be made available to the community upon submission of this article, alongside other value-added data associated with MWS (full details are given in the data availability statement).
The structure of these files is described in Appendix \ref{mock data model}. 

The AuriDESI mock catalogues can be used to test methods for inferring fundamental properties of the Milky Way thick disc and stellar halo from the DESI data. Potential applications include searches for structure in the stellar halo, studies of the properties and dynamical influence of satellites and their streams, radial trends in stellar populations and methods to reconstruct the assembly history of the Galaxy. The sample of 6 haloes, all with some resemblance to the real Milky Way, allows investigations of how halo-to-halo variations in accretion history may affect the observable properties of stellar haloes (and hence tests of methods to distinguish between different accretion histories()). Likewise, the four alternative solar positions for each halo can be used to assess the impact of having only a partial view of the Galaxy. However, we find that, with a DESI-like footprint and selection function, changing the solar position does not produce significant differences in the large-scale content of the resulting mock surveys. Although there are small differences in the total stellar density profile of sparse tracers at large distances, the same set of ten most-massive progenitors is visible  in each halo when viewed from different angular positions in the disc, with only small differences in the rank order of the least massive. We conclude that a survey on the scale of MWS provides a good sampling of the bulk of the mass in a Milky Way-like stellar halo.

AuriDESI is an attempt to provide realistic mock data sets for the next generation of wide-area, densely sampled spectroscopic surveys. DESI MWS will soon be joined by 4MOST \citep{4most}, WEAVE \citep{Weave}, SDSS-V \citep{SDSSV} and PFS \citep{PFS2022}. In view of the need to extend and improve the techniques for making catalogues that address all these surveys, we briefly discuss several important caveats to the methods used to produce AuriDESI.

First, on the technical side, the ICC-MOCKS method uses the \enbid{} algorithm to estimate the phase-space volumes associated with simulation star particles, over which their associated stars are distributed. This is a form of adaptive kernel smoothing, in which the kernel bandwidth in each dimension is proportional to the corresponding side-length of the hyper-cubic volume estimated by \enbid{}. As described in \citet{Lowing2015}, this estimate of phase-space volume is necessarily approximate. 
When the local (six-dimensional) phase space is sparsely sampled by the original star particles, the algorithm results in smoothed stellar distributions that are noisy near the kernel bandwidth. The consequence is an artificial clumping of stars in the catalogues. These clumping artefacts needs to be kept in mind when using the catalogues to examine phase-space distributions at smaller scales. The scaling of the kernel bandwidth, and the shape of the kernel (by default a hyper-ellipsoid) are also somewhat arbitrary. It is likely that other, more accurate techniques for smoothing the simulated stellar mass distribution can be developed (although any smoothing is inevitably less accurate than a simulation with higher resolution). \citet{SHLim2022} discuss a technique called \galaxyflow{}, which they demonstrate can improve on \enbid{}-based kernel smoothing methods in certain applications. The applications they describe are currently limited to the extended solar neighbourhood. Further technical improvement is necessary to apply this method efficiently to N-body simulations over the radial range probed by DESI.

Second, other limitations mentioned in \citet{Grand2018} concerning the AuriGaia ICC mocks also apply to AuriDESI. For instance, although the phase space density estimate uses merger tree information to avoid cross-talk between stars associated with different accreted progenitors, all star particles in the \insitu{} component are considered as potential phase-space neighbours of one another. This is a limitation, because the \insitu{} component most likely comprises several different populations in phase space (at minimum, an \insitu{} disc and stellar halo). In more detail, our present treatment potentially smooths over distinct dynamical structures in the disc (although this is a concern, we note that distinct spiral arms traced by young stars are still seen in AuriGaia). 

Third, although the Auriga simulations have been shown to be good representations of Milky Way-like galaxies in many respects \citep{Grand2017}, there are some important ways in which they do not resemble the real Milky Way. In particular, the accreted stellar haloes are too metal-rich and likely too massive. As discussed above, such discrepancies arise both from the sub-grid models used in the simulation and the criteria used to construct samples of Milky Way analogues. Future suites of simulations may define samples based on criteria other than host halo mass: for example, analogues of well-known features of the Milky Way such as the LMC or a GSE-like accretion event. This would provide a denser sampling of accretion histories resembling that of the Milky Way.

Finally, mismatches between the mocks and the Milky Way in the density of sources belonging to the different MWS target classes has an important effect on the spectroscopic mock catalogues. A \mainblue{} or \mainred{} star has higher priority in fiber assignment, relative to a \mainbroad{} or \texttt{MWS-FAINT} star. Since the Auriga stellar haloes are more massive than the stellar halo of the MW, there are typically 10 times more \mainblue{} and \mainred{} stars in the the AuriDESI photometric target catalogues than in the real DESI photometric target catalogue. When fiber assignment is carried out, this much larger pool of higher-priority \mainblue{} or \mainred{} targets consumes a much larger fraction of the fiber budget, resulting in a spectroscopic catalogue with a much lower fraction of \mainbroad{} and \texttt{MWS-FAINT} stars compared to the real survey. This is a faithful representation of what the DESI survey would observe in the Auriga galaxies (given the same total number of fibers opportunities), but it may limit comparisons to these lower-priority classes. For example, the larger number of intrinsically bright, low proper-motion metal-rich halo stars (\mainred{}) in the AuriDESI photometric target catalogues means that very few intrinsically faint, high-proper motion, metal-rich (\mainbroad{}) disc stars receive mock fibers, which makes AuriDESI potentially somewhat less useful for the study of disc kinematics (this can explored using the `resampled' variants of the spectroscopic mock catalogues, described in Section~\ref{sec:resampled_mocks}).

We intend to address the technical limitations of these mocks in future work, and to provide updated versions of AuriDESI alongside future MWS data releases. The method can be readily applied to new simulations, including variants assuming different phenomenology for the dark matter. 


\section*{Acknowledgements}

The authors thank the anonymous referee, for comments that greatly improved the structure and clarity of the paper, and John Helly, for assistance with the AuriGaia mock catalogues and Auriga data products. NK and APC thank Alis Deason, David Weinberg and Risa Wechsler for their valuable comments. NK and APC acknowledge support from a Taiwan Ministry of Education Yushan Fellowship awarded to APC and the Taiwan National Science and Technology Council grants 109-2112-M-007-011-MY3 and 112-2112-M-007-017-MY3. This work used high-performance computing facilities operated by the Center for Informatics and Computation in Astronomy (CICA) at National Tsing Hua University. This equipment was funded by the Taiwan Ministry of Education, the Taiwan National Science and Technology Council, and National Tsing Hua University. AHR is supported by a Research Fellowship from the Royal Commission for the Exhibition of 1851. SK acknowledges support from Science \& Technology Facilities Council (STFC) (grant ST/Y001001/1). RG is supported by an STFC Ernest Rutherford Fellowship (ST/W003643/1).

This material is based upon work supported by the U.S. Department of Energy (DOE), Office of Science, Office of High-Energy Physics, under Contract No. DE–AC02–05CH11231, and by the National Energy Research Scientific Computing Center, a DOE Office of Science User Facility under the same contract. Additional support for DESI was provided by the U.S. National Science Foundation (NSF), Division of Astronomical Sciences under Contract No. AST-0950945 to the NSF’s National Optical-Infrared Astronomy Research Laboratory; the Science and Technology Facilities Council of the United Kingdom; the Gordon and Betty Moore Foundation; the Heising-Simons Foundation; the French Alternative Energies and Atomic Energy Commission (CEA); the National Council of Science and Technology of Mexico (CONACYT); the Ministry of Science and Innovation of Spain (MICINN), and by the DESI Member Institutions: \url{https://www.desi.lbl.gov/collaborating-institutions}. Any opinions, findings, and conclusions or recommendations expressed in this material are those of the author(s) and do not necessarily reflect the views of the U. S. National Science Foundation, the U. S. Department of Energy, or any of the listed funding agencies.

The authors are honored to be permitted to conduct scientific research on Iolkam Du’ag (Kitt Peak), a mountain with particular significance to the Tohono O’odham Nation.

For the purpose of open access, the author has applied a Creative Commons Attribution (CC BY) licence to any Author Accepted Manuscript version arising from this submission.

This work made use of the following software: NumPy \citep{numpy}, SciPy \citep{scipy}, Astropy \citep{astropy:2013,astropy:2018}, Matplotlib \citep{matplotlib}, Healpy \citep{healpy,healpy1}, PyGaia (A. Brown; Gaia Project Scientist Support Team and the Gaia DPAC; \url{https://github.com/agabrown/PyGaia}). The ICC-mock processing pipelines used in this paper were originally written by Ben Lowing, with further contributions by John Helly and APC.

\section*{Data Availability}

All data shown in the figures, or necessary to reproduce them, are available in a machine-readable form on Zenodo \url{https://doi.org/10.5281/zenodo.10553709}. The mock catalogues themselves will be made available upon publication, along with documentation, at \url{https://data.desi.lbl.gov/public/papers/mws/auridesi/v1}.



\bibliographystyle{mnras}
\bibliography{bibliography} 




\appendix
\section{\mwsmain{} and \texttt{MWS-FAINT} sample selection criteria}
\label{name:mws selection criteria}
This section describes the exact criteria used to select stars in the MWS \mainblue{}, \mainred{} and \mainbroad{} target categories. Identical criteria are used to generate the real MWS photometric target catalogue from the combination of LS and Gaia photometry, and to generate the AuriDESI photometric target catalog from AuriGaia, extended with LS photometry as described in the text. The only exception is the Gaia data column \texttt{astrometric\_params\_solved} which, for the real Gaia data, indicates if the position and astrometric quantities are solved for a particular source). This criterion is used only when generating the real MWS photometric target catalogue.

In the following definitions, $f_{r}$ is the $r$-band flux and $f_\mathrm{r,19} = 10^{(22.5-19)/2.5}$.
\begin{itemize}
    \item \mwsmain{}: $16 < r < 19$.  All targets in this magnitude range are classified into one of the following mutually exclusive subsets:

\begin{itemize}
    \item \mainblue{}:
    
    \begin{enumerate}
        \item $g - r < 0.7$.
    \end{enumerate}
    
    \item \mainred{}: 
        \begin{enumerate}
            \item $g - r > 0.7$;
            \item $\pi < (3\sigma_\pi +0.3)\,\mathrm{mas}$ AND $|\mu| < 5 \sqrt{ f_{r} / f_\mathrm{r,19}} \,\mathrm{mas\,yr^{-1}}$.
            \item \texttt{astrometric\_params\_solved} $\geq 31$
    \end{enumerate}
    \item \mainbroad{}: 
    \begin{enumerate}
        \item  $g - r > 0.7$;
         \item $\pi > (3\sigma_\pi +0.3)\,\mathrm{mas}$ OR $|\mu| > 5 \sqrt{ f_{r} / f_\mathrm{r,19}} \,\mathrm{mas\,yr^{-1}}$.
         \item \texttt{astrometric\_params\_solved} $<31$
    \end{enumerate}
    \end{itemize}
\end{itemize}

\begin{itemize}
    \item \texttt{MWS-FAINT}: $19 < r < 20$. All targets in this magnitude range are classified into one of the following mutually exclusive subsets:

    \begin{itemize}
    \item \texttt{FAINT-BLUE}: 
    \begin{enumerate}
    \item  $g - r < 0.7$.
    \end{enumerate}
    \item \texttt{FAINT-RED}:
    \begin{enumerate}
        \item $g - r > 0.7$ 
        \item $\pi < (3\sigma_\pi +0.3)\,\mathrm{mas}$ AND $|\mu| < 3 \,\mathrm{mas\,yr^{-1}}$.
        \item \texttt{astrometric\_params\_solved} $\geq 31$
    \end{enumerate}
    \end{itemize}
\end{itemize}

There is no equivalent of the \texttt{MAIN-BROAD} subset in the faint sample.
 
In the AuriDESI mocks, the categories to which each star belongs are encoded in the \texttt{MWS\_TARGET} bitmask. For example, the mask \texttt{(MWS\_TARGET \& MAIN\_RED) != 0} selects \texttt{MAIN-RED} targets in the mock. For more information on the use of these bitmasks, see \citet{myers22a}.

\section{AuriDESI isochrones}
\label{isochrones}

As mentioned in Sections~\ref{sec:star_particle_expansion} and \ref{sec:footprint_and_targets}, when computing stellar parameters for each star, the initial mass of the parent star particle is interpolated onto the nearest age and metallicity point of the available isochrone grid. 

\begin{figure}
    \centering
    \includegraphics[width=\linewidth]{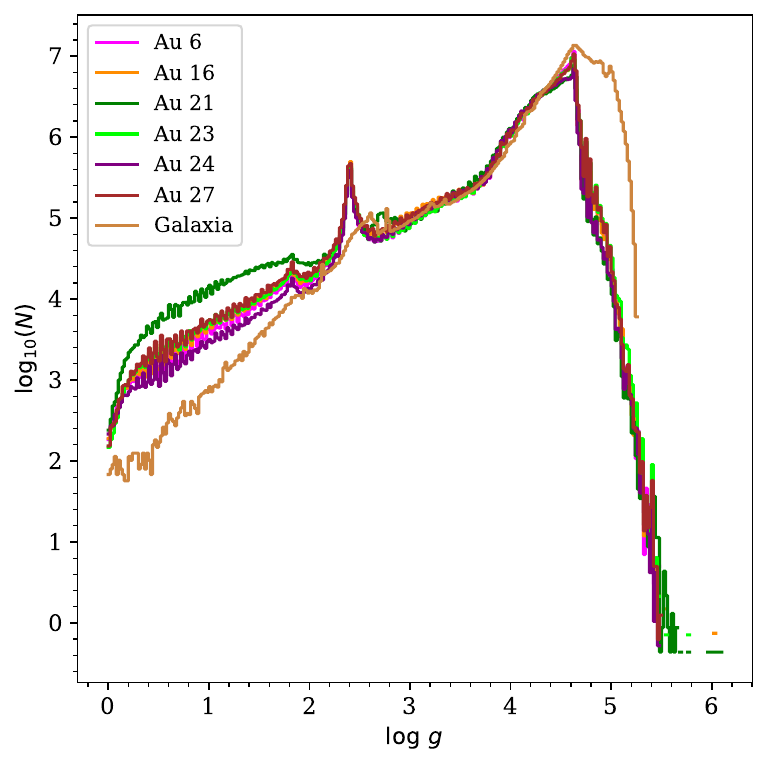}
    \caption{Surface gravity ($\log g$) distributions for all stars in the mock MWS footprint, in each of the six AuriGaia haloes. The distributions are normalised to have the same integral over the range $3 < \log g <6$. For comparison, the light brown line shows the default Galaxia model from \citet{Sharma2011} modified to have a broken power-law halo with a steeper outer slope that better approximates recent Milky Way results \citep[see][]{2022APC}.}
    \label{fig:logghist}
\end{figure}

Fig. \ref{fig:logghist}, shows the resulting $\log g$ distributions of stars in the MWS footprint, comparing the AuriGaia haloes to the Galaxia model. It is quite clear that there a deficit of stars with $\log g > 4.5$ in AuriGaia relative to Galaxia. This lack of low mass stars (associated with the thin disc of the galaxy) arises from the difference in the isochrones used by Galaxia and AuriGaia. The ICC mocks of AuriGaia use version 1.2S of the PARSEC isochrones \citep{Bressan2012,Chen2014,Chen2015,Tang2014,Grand2018}. Galaxia uses an older version of the PARSEC models \citep[generated from CMD v2.1 in 2008,][]{Marigo2008,Marigo_Girardi2007,Sharma2011}. In the more recent PARSEC models, fainter absolute magnitudes are assigned to low mass stars. This change is described by \citet{Chen2014}. We intend to address this apparent discrepancy in future versions of AuriDESI. For this reason, we caution against the use of the mock catalogues we provide here for applications that are sensitive to the density of low-mass stars in the nearby thin disc. 

The differences between the distributions at high $\log g$ can be attributed to the halo-to-halo variation in the slope of the outer halo  density profile. The significant excess  in the $\log g$ distribution for Au 21 is due to the contribution of a single satellite (see \ref{Satellites}). 

\section{Mock catalogue data model}
\label{mock data model}

This section describes the data model of the AuriDESI mock catalogues. We provide each catalog as a FITS file with five extensions. These follow essentially the same data model as the real MWS catalog data products \citep{Koposov2024}. Table \ref{tab:RVTAB} describes the RVTAB extension, which includes stellar parameters such as radial velocity, effective temperature, metallicity and their corresponding uncertainties. In the real MWS survey, these will be obtained from reductions of the observed spectra with the MWS analysis pipelines. Table \ref{tab:Fibermap} describes the FIBERMAP extension. This includes the sky positions and other targeting information for each star. Table \ref{tab:Gaia} provides Gaia observables, as in AuriGaia but here updated to the EDR3 error model used for DESI targeting. Tables \ref{tab:true} and \ref{tab:Columns in PROGENITOR extension in AuriDESI} contain information propagated from the underlying Auriga simulation: respectively, the original ("true") simulated values of various quantities for the parent star particle of each star, and information concerning the bulk properties and orbital history of each accreted progenitor.

\begin{table*}
    \centering
    \begin{tabular}{|l|l|l|}
           \hline
        Name & Description & Units  \\
        \hline
         VRAD & Radial velocity & km/s \\
         VRAD\_ERR &  Error in radial velocity & km/s\\
         LOGG & Logarithm of surface gravity & dex\\
         TEFF & Effective temperature & K\\
         ALPHAFE & Alpha abundance & dex\\
         FEH & Metallicity & dex \\
         LOGG\_ERR & Error in surface gravity & dex\\
         TEFF\_ERR & Error in effective temperature & K\\
         ALPHAFE\_ERR & Error in alpha abundance & dex\\
         FEH\_ERR & Error in metallicity & dex \\
         TARGET\_RA & Right ascension  & degrees\\
         TARGET\_DEC & Declination  & degrees \\
         TARGETID & Unique identifier for a star & \\
         HEALPIX & Healpix value corresponding to sky positions\\
         SURVEY & DESI Survey program (Main/SV) \\
         PROGRAM & DESI observing program (Bright/Dark)\\
         \\
           \hline
    \end{tabular}
    \caption{Columns in the RVTAB extension of an AuriDESI catalogue file. }
    \label{tab:RVTAB}
\end{table*}

\begin{table*}
    \centering
    \begin{tabular}{|l|l|l|}
           \hline
            Name & Description & Units   \\
            \hline
             TARGETID & Unique identifier for a star &  -- \\
             TARGET\_RA & Right ascension  & degrees\\
             TARGET\_DEC & Declination & degrees \\
             PMRA & Proper motion right ascension & mas/yr\\
             PMDEC & Proper motion declination & mas/yr\\
             FATYPE & Class of the star for fiber assignment\\
             OBJTYPE & Type of the object & --   \\
             MORPHTYPE & Morphological type & --  \\
             FLUX\_G & Flux in DECam $g$ band &nanomaggies\\
             FLUX\_R & Flux in DECam $r$ band &nanomaggies\\
             FLUX\_Z & Flux in DECam $z$ band &nanomaggies\\
             GAIA\_PHOT\_G\_MEAN\_MAG & Gaia G mag & mag\\
             GAIA\_PHOT\_BP\_MEAN\_MAG & Gaia BP mag & mag\\
             GAIA\_PHOT\_RP\_MEAN\_MAG & Gaia RP mag & mag\\
             PARALLAX & Parallax  & mas \\
             PHOTOSYS & Photometric system & -- \\
             DESI\_TARGET & Target class of DESI the star belong to & --  \\
             MWS\_TARGET & Is the star a MWS target & --\\
             SV1\_MWS\_TARGET & does the star come within DESI SV1 field & -- \\
             SV2\_MWS\_TARGET & does the star come within DESI SV2 field  & -- \\
             SV3\_MWS\_TARGET & does the star come within DESI SV3 field  & -- \\
           \hline
    \end{tabular}
    \caption{Columns in the \texttt{FIBERMAP} extension of an AuriDESI catalogue file.}
    \label{tab:Fibermap}
\end{table*}

\begin{table*}
    \centering
    \begin{tabular}{|l|l|l|}
           \hline
            Name & Description & Units   \\
            \hline
             RA & Right ascension  & degrees\\
             RA\_ERROR & Error in right ascension  & degrees\\
             DEC & Declination& degrees \\
             DEC\_ERROR & Error in declination  & degrees \\
             PARALLAX & Parallax  & mas \\
             PARALLAX\_ERROR & Error in parallax  & mas \\
             PM & Proper motion  & mas \\
             PMRA & Proper motion in  right ascension & mas/yr\\
             PMRA\_ERROR & Error in proper motion in right ascension & mas/yr\\
             PMDEC & Proper motion in declination & mas/yr\\
             PMDEC\_ERROR & Error in proper motion in declination & mas/yr\\
             RADIAL\_VELOCITY & Radial velocity & km/s \\
             RADIAL\_VELOCITY\_ERROR &  Error in radial velocity & km/s\\ 
            
           \hline
    \end{tabular}
    \caption{Columns in the GAIA extension of an AuriDESI catalogue file.}
    \label{tab:Gaia}
\end{table*}

\begin{table*}
    \centering
    \begin{tabular}{|l|p{12cm}|l|}
           \hline
        Name & Description & Units  \\
        \hline
         PARTICLE\_ID & Particle ID of the parent star particle & -- \\
         RA & Right ascension of the star & degrees\\ 
         DEC &  Declination of the star & degrees \\
         PARALLAX & Parallax of the star & mas \\
         PMRA & Proper motion right ascension & mas/yr\\
         PMDEC & Proper motion declination & mas/yr\\
         VRAD & Radial velocity & km/s \\
         X\_PARENT & Galactocentric X coordinate of the parent star particle & kpc \\
         Y\_PARENT & Galactocentric Y coordinate of the parent star particle & kpc \\
         Z\_PARENT & Galactocentric Z coordinate of the parent star particle & kpc \\
         VX\_PARENT & Velocity of parent star particle along galactocentric coordinate X & km/s \\
         VY\_PARENT & Velocity of parent star particle along galactocentric coordinate Y & km/s \\
         VZ\_PARENT & Velocity of parent star particle along galactocentric coordinate Z & km/s \\
         AGE & Age of the star & Gyr\\
         TEFF & Effective temperature of the star & K\\
         LOGG & Surface gravity of the star & dex\\
         Z & Metallicity of the star & dex\\
         FEH & Iron abundance of the star & dex\\
         ALPHAFE & Alpha abundance of the star & dex\\
         MASS & Mass of the star & $\msol$\\
         INITIAL\_MASS & Initial mass of the star & $\msol$\\
         GRAVPOTENTIAL & Gravitational potential energy of the parent star particle & km$^2/s^2$ \\
         APP\_GMAG & Apparent magnitude in DECAM $g$ - band & mag\\
         APP\_RMAG  & Apparent magnitude in DECAM $r$ - band & mag\\
         APP\_ZMAG & Apparent magnitude in DECAM $z$ - band & mag\\
         POP\_ID & 0: star formed \insitu{}; 1: star was accreted; 2: star is still bound to a satellite & -- \\
         TREE\_ID & Integer label for each unique progenitor system. 0: star formed \insitu{} (i.e. in the main branch of the merger tree of the MW analogue); $>0$: star formed in an accreted satellite. & -- \\
         SUBHALO\_NR & Integer label for the present-day subhalo to which the star is currently bound. 0: star bound to main halo; $>0$: star is bound to a surviving satellite & -- \\
           \hline
    \end{tabular}
    \caption{Columns in the TRUE\_VALUE extension of an AuriDESI catalogue.}
    \label{tab:true}
\end{table*}
\begin{table*}
    \centering
    \begin{tabular}{|l|l|l|}
           \hline
        Name & Description & Units  \\
        \hline
        TREE\_ID & Identifier of progenitor, corresponding to a specific merger tree at redshift 0 &  -- \\
        SUBHALO\_NR & subhalo the star particle belong to redshift 0 &  -- \\
        MSTAR\_TOT & Total mass of star particles with a given TREE\_ID & $\msol$ \\
        MSTAR\_ACCRETED\_MAIN & Stellar mass for a given TREE\_ID associated with the main halo (SUBHALO\_NR=0) at z=0 &  $\msol$ \\
        MSTAR\_FUZZ & Stellar mass for a given TREE\_ID associated with the fuzz (SUBHALO\_NR=-1) at z=0 & $\msol$ \\
        HAS\_BOUND\_SAT & Does the subhalo with a given TREE\_ID contain particles at z =0 & -- \\
        MSTAR\_SAT & Stellar mass for a given TREE\_ID associated with any SUBHALO\_NR > 0 at z=0 & $\msol$ \\
        Z\_INFALL & Infall redshift for a given TREE\_ID & -- \\
        HALO\_MSTAR\_INFALL & Total stellar mass of the halo specified in the latest Auriga merger trees & $\msol$ \\
        HALO\_MGAS\_INFALL & Total gas mass of the halo specified in the latest Auriga merger trees & $\msol$\\
        HALO\_MDM\_INFALL & Total dark matter mass of the halo specified in the latest Auriga merger trees & $\msol$ \\
        SUBHALO\_MSTAR & SUBFIND stellar mass for subhalo for a given TREE\_ID at redshift 0 & $\msol$ \\
        SUBHALO\_MGAS & SUBFIND gas mass for subhalo for a given TREE\_ID at redshift 0 & $\msol$ \\
        SUBHALO\_MDM & SUBFIND dark matter mass for subhalo for a given TREE\_ID at redshift 0 & $\msol$\\
        R\_GAL & Present day galctocentric distance (-1 if no surviving subhalo) & kpc \\
         \hline
    \end{tabular}
    \caption{Columns in PROGENITOR extension of AuriDESI catalogue.}
    \label{tab:Columns in PROGENITOR extension in AuriDESI}
\end{table*}

\section{Au 21 Progenitors}
\label{halo21_appendix}
To illustrate variations between the different Auriga halos, we repeat some of the plots shown for Au 6 in section \ref{Halo 6} in the context of Au 21.

\begin{figure*}
    \centering
    \includegraphics[width=15cm]{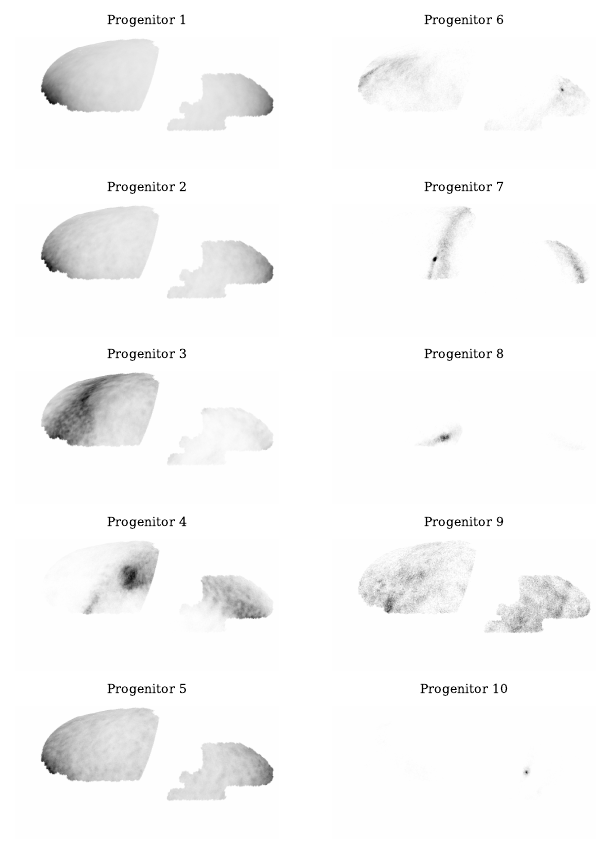}
    \caption{On-sky density distribution of stars from 10 most massive progenitors in the Au 21 AuriDESI mock catalogue. The densest regions typically correspond to the position of the remnant of the main body of the progenitor satellite.}
    \label{fig:H21_Progenitors}
\end{figure*}

Fig. \ref{fig:H21_Progenitors} shows the projected density of stars from the ten most massive progenitors in Au 21 over the DESI footprint. Some of these progenitors show clear stream-like features (progenitors 3, 4, 7, and 8) while others are distributed smoothly (progenitors 1, 2, and 5). Most of these progenitors (3, 4, 6, 7, 8, and 10) are associated with a region of high density, which indicates the positions of their satellite remnants (see Section \ref{Satellites}). The clumpiness seen in these sky distributions is a result of kernel smoothing algorithm, as described in the main text.

\begin{figure*}
    \centering
    \includegraphics[width=\linewidth]{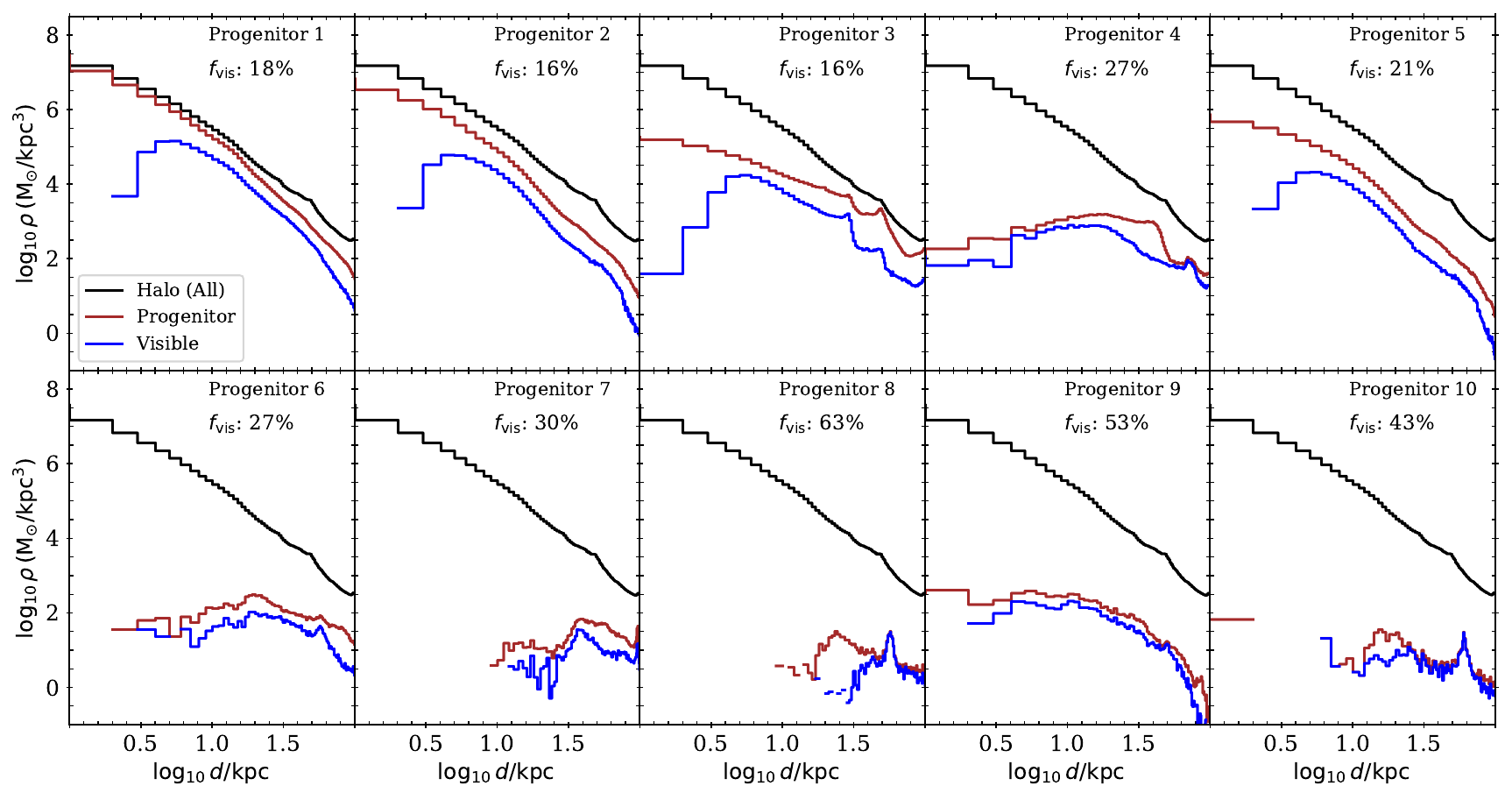}
    \caption{Galactocentric density distribution of star particles in the 10 most massive progenitors of Au 21. The black line shows the density profile of all the accreted star particles in the halo (identical in each panel). The brown line shows the profile of all the star particles from a given progenitor and the blue line shows the profile of all star particles that contribute at least one star to AuriDESI mock catalogue. The $f_\mathrm{vis}$ value in each panel is the fraction of mass visible to DESI.}
    \label{fig:density_h21_progenitors}
\end{figure*}

Fig. \ref{fig:density_h21_progenitors} shows the radial density profiles of simulated star particles in the 10 most massive progenitors of Au 21. In each panel, the black line shows the density profile of all the accreted star particles in Au 21, the brown line shows the profile of all the star particles in a given progenitor, and the blue line shows the profile for only those star particles that contribute at least one star to the AuriDESI mock catalogue. The brown and blue lines are similar, except at lower galactocentric distances. This difference comes mainly from the geometric effect of the DESI footprint, which is restricted to high latitudes (|b| > 20).

\begin{figure*}
    \centering
    \includegraphics[width=\linewidth]{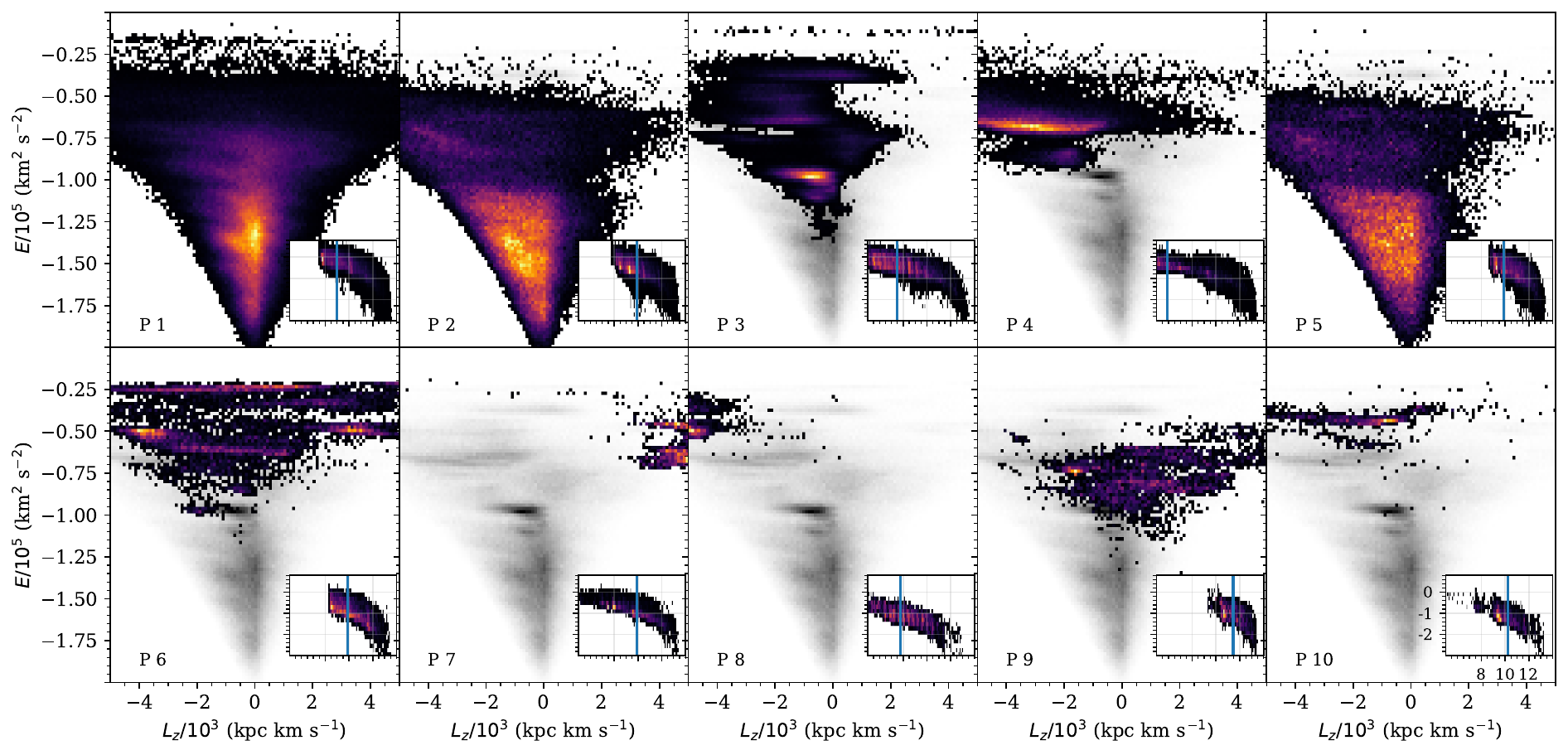}
    \caption{Energy - angular momentum distributions of accreted star particles from the 10 most massive progenitors of Au 21. The grey scale distribution in the background corresponds to the distribution of all star particles from all the progenitors. The inset panels show the distribution of stellar ages (Gyr, horizontal axis) and metallicities (dex, vertical axis). All inset panels have the same range of age and metallicity (the scale of each axis is shown on the last panel, and omitted in the others).}
    \label{fig:E_Lz_h21_progenitors}
\end{figure*}

Fig.~\ref{fig:E_Lz_h21_progenitors} shows the distribution of star particles from the ten most massive progenitors of Au 21 in the space of orbital energy and angular momentum. The inset plots in each panel show the age-metallicity distribution of the same particles. These plots provide an outline of the dynamical history of each progenitor. For example, progenitors 1, 2, and 5 are distributed over a very broad range of energy; 2 and 5 have a more prograde rotation while 1 seems to be more relaxed, with no bulk rotation. Star formation in progenitors 1, 2, 5, 6, 9, and 10 extends up to $\sim$ 8 Gyr, while the others sustain star formation up to $\sim$ 6 Gyr before the present.

\section*{Author affiliations}

\textit{
$^{1}$Institute of Astronomy and Department of Physics, National Tsing Hua University, Hsinchu 30013, Taiwan\\
$^{2}$Center for Informatics and Computation in Astronomy, National Tsing Hua University, Hsinchu 30013, Taiwan\\
$^{3}$Institute for Computational Cosmology, Department of Physics, Durham University, South Road, Durham DH1 3LE, UK\\
$^{4}$Institute for Astronomy, University of Edinburgh, Royal Observatory, Blackford Hill, Edinburgh EH9 3HJ, UK\\
$^{5}$Institute of Astronomy, University of Cambridge, Madingley Road, Cambridge CB3 0HA, UK\\
$^{6}$Kavli Institute for Cosmology, University of Cambridge, Madingley Road, Cambridge CB3 0HA, UK\\
$^{7}$Lawrence Berkeley National Laboratory, 1 Cyclotron Road, Berkeley, CA 94720, USA\\
$^{8}$Physics Dept., Boston University, 590 Commonwealth Avenue, Boston, MA 02215, USA\\
$^{9}$Instituto de Astrof\'{i}sica de Canarias, C/ Vía L\'{a}ctea, s/n, E-38205 La Laguna, Tenerife, Spain \\ $^{10}$Universidad de La Laguna, Dept. de Astrof\'{\i}sica, E-38206 La Laguna, Tenerife, Spain\\
$^{11}$Department of Physics \& Astronomy, University College London, Gower Street, London, WC1E 6BT, UK\\
$^{12}$Department of Physics and Astronomy, The University of Utah, 115 South 1400 East, Salt Lake City, UT 84112, USA\\
$^{13}$Instituto de F\'{\i}sica, Universidad Nacional Aut\'{o}noma de M\'{e}xico,  Cd. de M\'{e}xico  C.P. 04510,  M\'{e}xico\\
$^{14}$Departamento de F\'isica, Universidad de los Andes, Cra. 1 No. 18A-10, Edificio Ip, CP 111711, Bogot\'a, Colombia\\
$^{15}$Observatorio Astron\'omico, Universidad de los Andes, Cra. 1 No. 18A-10, Edificio H, CP 111711 Bogot\'a, Colombia\\
$^{16}$Institut d'Estudis Espacials de Catalunya (IEEC), 08034 Barcelona, Spain\\
$^{17}$Institute of Cosmology \& Gravitation, University of Portsmouth, Dennis Sciama Building, Portsmouth, PO1 3FX, UK\\
$^{18}$Institute of Space Sciences, ICE-CSIC, Campus UAB, Carrer de Can Magrans s/n, 08913 Bellaterra, Barcelona, Spain\\
$^{19}$Department of Astronomy, University of Michigan, Ann Arbor, MI 48109, USA \\
$^{20}$Astrophysics Research Institute, Liverpool John Moores University, 146 Brownlow Hill, Liverpool, L3 5RF, UK\\ 
$^{21}$Center for Cosmology and AstroParticle Physics, The Ohio State University, 191 West Woodruff Avenue, Columbus, OH 43210, USA\\
$^{22}$Department of Physics, The Ohio State University, 191 West Woodruff Avenue, Columbus, OH 43210, USA\\
$^{23}$Department of Physics, Southern Methodist University, 3215 Daniel Avenue, Dallas, TX 75275, USA\\
$^{24}$Institut de F\'{i}sica d’Altes Energies (IFAE), The Barcelona Institute of Science and Technology, Campus UAB, 08193 Bellaterra Barcelona, Spain\\
$^{25}$Departament de F\'{i}sica, Serra H\'{u}nter, Universitat Aut\`{o}noma de Barcelona, 08193 Bellaterra (Barcelona), Spain\\
$^{26}$NSF's NOIRLab, 950 N. Cherry Ave., Tucson, AZ 85719, USA\\
$^{27}$Instituci\'{o} Catalana de Recerca i Estudis Avan\c{c}ats, Passeig de Llu\'{\i}s Companys, 23, 08010 Barcelona, Spain\\
$^{28}$National Astronomical Observatories, Chinese Academy of Sciences, A20 Datun Rd., Chaoyang District, Beijing, 100012, P.R. China\\
$^{29}$ Instituto de Astrof\'{i}sica de Andaluc\'{i}a (CSIC), Glorieta de la Astronom\'{i}a, s/n, E-18008 Granada, Spain\\
$^{30}$Department of Physics, Kansas State University, 116 Cardwell Hall, Manhattan, KS 66506, USA\\
$^{31}$Department of Physics and Astronomy, Sejong University, Seoul, 143-747, Korea\\
$^{32}$CIEMAT, Avenida Complutense 40, E-28040 Madrid, Spain\\
$^{33}$Department of Physics, University of Michigan, Ann Arbor, MI 48109, USA\\
$^{34}$Department of Physics \& Astronomy, Ohio University, Athens, OH 45701, USA}

\bsp	
\label{lastpage}
\end{document}